\documentclass[aps,pra,reprint,nofootinbib,floatfix]{revtex4-2}
\usepackage{amsmath,amssymb,bm}
\usepackage{graphicx}
\usepackage[colorlinks=true,linkcolor=blue,citecolor=blue,urlcolor=blue]{hyperref}
\usepackage{microtype}
\usepackage{orcidlink}

\newcommand{\Irr}{\mathfrak I}
\newcommand{\Tr}{\operatorname{Tr}}
\newcommand{\ket}[1]{|#1\rangle}
\newcommand{\bra}[1]{\langle #1|}
\newcommand{\proj}[1]{|#1\rangle\!\langle #1|}
\newcommand{\id}{\mathbb{I}}
\newcommand{\cS}{\mathcal S}
\newcommand{\cA}{\mathcal A}
\newcommand{\cB}{\mathcal B}
\newcommand{\cC}{\mathcal C}
\newcommand{\Info}{\mathcal I}

\begin{document}

\title{Quantifying the irrealism required by quantum random access codes}

\author{Pedro R. Dieguez\orcidlink{0000-0002-8286-2645}}
\email{pedrodieguez@ufpr.br}
\affiliation{Department of Physics, Federal University of Paran\'{a}, 81531-980 Curitiba, Paran\'{a}, Brazil}

\begin{abstract}
How far must decoding observables be from having elements of reality to achieve a given quantum advantage in a communication task? We answer this question for quantum random access codes. Here, an observable has an element of reality when its unrevealed measurement leaves the preparation undisturbed; its irrealism is the relative entropy to the closest such preparation. Because it comprises local coherence and the correlations destroyed by the unrevealed measurement, irrealism applies to prepare-and-measure (PM) and entanglement-assisted (EA) codes alike. For PM codes decoded in mutually unbiased bases, we obtain the irrealism frontier, the least total irrealism compatible with each success probability, exactly in every dimension; an excess-irrealism identity shows that Gibbs states of the decoding operator attain it uniquely. For PM codes minimized also over the decoders, the required irrealism vanishes up to the classical value and is positive beyond it; for qubits, mutually unbiased decoders are not the cheapest below a success probability of $0.833$. For EA codes with a qubit receiver, trusted decoders, and parity-dependent sender settings, the frontier exceeds the PM one by the realism-based nonlocality of the contexts used by an optimal code, and with rank-one projective sender measurements a success probability above $0.815$ certifies a minimum realism-based nonlocality of the shared state.
\end{abstract}

\maketitle

\section{Introduction}

Does a physical quantity have a value before anyone looks? Einstein, Podolsky, and Rosen (EPR) turned this question into a criterion: if, without disturbing a system, one can predict with certainty the value of a quantity, then there is an element of physical reality corresponding to it~\cite{EPR1935}. Bohr's reply moved the question from the system to the experiment, holding that the conditions which define a quantity cannot be separated from the arrangement used to observe it~\cite{Bohr1935}. The question has since acquired an operational edge. Quantum information processing usually draws its advantages from preparations in which the relevant quantities have no definite values, and it becomes natural to ask how much of this indefiniteness a given performance of a task requires.

Bilobran and Angelo proposed an operational criterion of realism that agrees with that of EPR for eigenstates and extends it to mixed preparations~\cite{Bilobran2015}. For a preparation $\rho_{\cA\cB}$ and a nondegenerate observable $B$ of the subsystem $\cB$, with projectors $\Pi_k=\proj{b_k}$, $B$ is an element of reality if a measurement of $B$ whose outcome is not revealed leaves the preparation unchanged,
\begin{equation}
\Phi_B(\rho_{\cA\cB})\equiv\sum_k(\id_\cA\otimes\Pi_k)\,\rho_{\cA\cB}\,(\id_\cA\otimes\Pi_k)=\rho_{\cA\cB} ,
\label{eq:criterion}
\end{equation}
that is, if $\rho_{\cA\cB}=\sum_kp_k\,\rho_\cA^{(k)}\otimes\proj{b_k}$ is a mixture of preparations in each of which $B$ has a definite value; a single system is the case of trivial $\cA$. The criterion concerns one observable at a time, and it differs from that of EPR: for a maximally entangled pair it assigns reality to neither of two complementary observables of one particle before the other particle is measured. Moreover, the criterion~\eqref{eq:criterion} asks whether a value is already encoded in the quantum preparation, not whether hidden-variable models exist. How far $B$ is from being an element of reality is measured by its irrealism
\begin{equation}
\begin{aligned}
\Irr_B(\rho_{\cA\cB})&\equiv\min_{\varrho}D\big(\rho_{\cA\cB}\Vert\Phi_B(\varrho_{\cA\cB})\big)\\
&= S\big(\Phi_B(\rho_{\cA\cB})\big)-S(\rho_{\cA\cB})\ge 0,
\end{aligned}
\label{eq:irr}
\end{equation}
where $S(\rho)=-\Tr(\rho\log_2\rho)$ is the von Neumann entropy and $D(\rho\Vert\sigma)=\Tr[\rho(\log_2\rho-\log_2\sigma)]$ the relative entropy, both in bits throughout. The irrealism is the relative-entropy distance from the preparation to the closest one in which $B$ is real, equals the increase in entropy produced by the unrevealed measurement, and vanishes exactly when $B$ is an element of reality. The measure has been applied to realism-based nonlocality~\cite{Gomes2018,Gomes2019,Fucci2019,gomes2022realism}, information--reality complementarity and the emergence of realism under monitoring~\cite{Dieguez2018,Orthey2025,10.1209/0295-5075/adc481,Lustosa2025}, the joint reality of incompatible observables~\cite{Caetano2024}, quantum-controlled interferometers~\cite{Dieguez2022}, and the Aharonov--Bohm effect~\cite{Paiva2023}. It has been incorporated into a resource theory~\cite{Costa2020}, axiomatized~\cite{OrtheyAngelo2022}, quantified by geometric monotones~\cite{PhysRevA.111.032212}, extended beyond quantum theory~\cite{fucci2024theory}, and measured in nuclear magnetic resonance~\cite{Dieguez2022} and photonic~\cite{Mancino2018,Lustosa2025} experiments.

For single-system preparations, irrealism reduces to the relative entropy of coherence in the eigenbasis of the decoding observable~\cite{Baumgratz2014,Streltsov2017,ChitambarGour2019}, whereas for bipartite preparations it also accounts for the correlations destroyed by the corresponding measurement~\cite{Bilobran2015}, and is therefore bounded from below by the local coherence plus the quantum discord~\cite{OllivierZurek2001,HendersonVedral2001}. This makes irrealism suited to quantum advantages in communication, since it captures both the advantage carried by the coherence of a transmitted quantum system and that enabled by shared entanglement, which coherence and correlation measures each capture only in part. The irrealism contains both contributions and separates them, and this separation is what relates the costs of the two architectures (Sec.~\ref{sec:EA}).
Indeed, in the context of quantum algorithms in which oracle problems are recast as communication tasks~\cite{Teeni2026}, one party encodes the oracle property to be learned into a quantum state and the other decodes it via measurement. There, the irrealism of the decoding measurement, which comprises the coherence of the measured state in the measurement basis and a discord-like term with the encoded property, vanishes precisely when both a coherence-based lower bound and the Holevo upper bound on the mutual information between the encoded property and the measurement outcome are saturated. Irrealism, therefore, enters directly into the performance of a communication task, but the amount of irrealism required to attain a given performance has not been determined.

\begin{figure*}[t]
\centering
\includegraphics[width=0.92\textwidth]{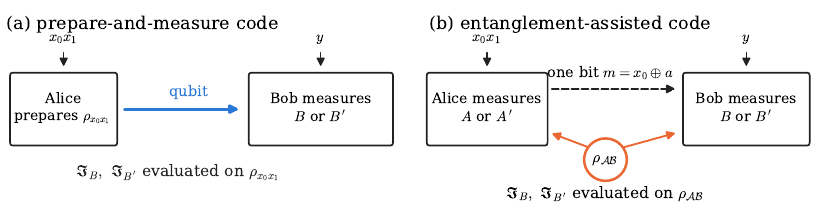}
\caption{The two architectures of the $2\to1$ QRAC. (a) Prepare-and-measure: Alice encodes $x_0x_1$ in a qubit; Bob, asked question $y$, measures $B$ to read $x_0$ or $B'$ to read $x_1$ ($B=\sigma_z$, $B'=\sigma_x$; in the notation of Sec.~\ref{sec:rac}, $(i,j)=(x_0,x_1)$). The irrealism of the decoding observables is evaluated on each transmitted state. (b) Entanglement-assisted~\cite{Pawlowski2010}: Alice measures her half of $\rho_{\cA\cB}$ with $A$ or $A'$, according to $x_0\oplus x_1$, and sends one bit; Bob measures $B$ or $B'$ on his half. The irrealism of Bob's decoding observables is evaluated on the shared state before Alice measures.}
\label{fig:schematic}
\end{figure*}

We determine it for quantum random access codes (QRACs)~\cite{Ambainis1998}. A random access code compresses several classical symbols into a single message of lower capacity, and its performance is the probability that the receiver recovers a symbol chosen after the encoding. Encoding into a quantum system~\cite{Ambainis1998}, or into a classical message assisted by shared entanglement~\cite{Pawlowski2010}, raises the probability of success above what any classical code allows, and QRACs have become a primitive of semi-device-independent cryptography~\cite{Pawlowski2011}, randomness generation~\cite{Li2011}, and self-testing~\cite{Farkas2019}. We consider codes in which two symbols are encoded, and Bob retrieves the requested one by measuring one of two decoding observables (Fig.~\ref{fig:schematic}). The advantage requires incompatible decoders~\cite{Carmeli2020} and, for any pair of decoders, states in which neither decoding observable is an element of reality (Secs.~\ref{sec:PM} and~\ref{sec:EA}). How far from reality the decoding observables must be is quantified by the irrealism frontier, the least total irrealism of the decoding observables compatible with a given success probability. Like the least entanglement compatible with a given Bell violation~\cite{Verstraete2002}, the frontier is a bound obeyed by every code rather than a strategy; with trusted decoders, it turns an observed success probability into a lower bound on the irrealism (Sec.~\ref{sec:gibbs}).

Our results are of three kinds. The first is the exact frontier for prepare-and-measure (PM) $2_d\to1$ codes~\cite{Tavakoli2015,Aguilar2018} decoded in mutually unbiased bases (MUBs), at every success probability and in every dimension (Sec.~\ref{sec:gibbs}). It is attained uniquely by Gibbs states of the decoding operator, and an upper bound obeyed by every code meets it at the quantum optimum, where the success probability alone fixes the irrealism, whatever the encoding. The proof rests on an identity of Pythagorean type, by which a closure property of the unrevealed measurements turns a stationary point into the unique global minimum; the identity applies to any weighted sum of irrealisms. We fix the decoders to MUBs because they are the only decoders that reach the quantum optimum~\cite{Farkas2019}. The second concerns the decoders themselves: minimized also over them, the least total irrealism vanishes exactly up to the classical value and is positive beyond it, in every dimension, and for qubits, where we obtain it exactly for every pair of decoders, MUBs are not the cheapest choice below a success probability $P_*\simeq0.833$ (Sec.~\ref{sec:decoders}). Known entropic relations enter as tools: the information--reality exclusion, the tripartite uncertainty relation with quantum memory~\cite{RenesBoileau2009,Berta2010} read in terms of irrealism, gives a lower bound that the exact frontier strictly exceeds (Sec.~\ref{sec:exclusion}). The third, described next, relates the two architectures.

For entanglement-assisted (EA) codes~\cite{Pawlowski2010} with a qubit receiver and trusted decoders, the bounds on each irrealism coincide with known device-independent and one-sided device-independent entropy bounds~\cite{Acin2007,Pironio2009,Branciard2012,Masini2022}. Building on them, we prove the closed-form least total irrealism for any quantum system on Alice's side with two-outcome measurements and show that it is attained; at the Tsirelson point it reaches two bits, with both decoding observables maximally unreal. In the optimal codes analyzed here, Bob's reduced state is maximally mixed, so his decoding observables have no local coherence and their irrealism lies entirely in the correlations with Alice. What appears as coherence in PM codes thus appears as correlations in EA codes. The EA frontier nonetheless lies above the PM one, because the EA irrealism is evaluated on the shared state, before Alice measures. Once she has measured, Bob's states conditioned on her outcome form a PM code, and relative to these states the decoding observables are more real. For rank-one projective measurements of Alice, this gain of reality is the realism-based nonlocality~\cite{Bilobran2015,Gomes2018} of the context formed by her observable and Bob's decoder. In the optimal code this gain, summed over the two decoders, is exactly the difference between the two frontiers, and for any code its average over the four contexts is bounded from below in terms of the success probability, so that a success probability above $0.815$ certifies a minimum realism-based nonlocality of the shared state (Sec.~\ref{sec:EA}).

Taken together, these results give the irrealism requirement three levels: beyond the classical value no decoding observable can be an element of reality; for mutually unbiased decoders each success probability requires a minimum irrealism; and at the quantum optimum the performance alone fixes the irrealism. Section~\ref{sec:reality} introduces the two architectures, the decomposition of irrealism, and the information--reality exclusion; Sec.~\ref{sec:cost} poses the question and determines the frontier for PM (Sec.~\ref{sec:PM}) and EA codes (Sec.~\ref{sec:EA}); Sec.~\ref{sec:conclusion} concludes.

\section{Random access codes and irrealism}\label{sec:reality}

\subsection{Random access codes}\label{sec:rac}

\emph{Prepare-and-measure codes.} In the $2_d\to1$ code~\cite{Tavakoli2015}, in which two $d$-valued symbols are encoded into one $d$-level system and one of them is retrieved, Alice receives $(i,j)\in\{0,\dots,d-1\}^2$ uniformly at random and sends a $d$-level state $\rho_{ij}$. Bob reads $i$ by measuring $B$, with eigenbasis $\{\ket{b_k}\}$ and projectors $\Pi_k=\proj{b_k}$, and $j$ by measuring $B'$, with eigenbasis $\{\ket{b'_k}\}$ and projectors $\Pi'_k$. In the standard code the two bases are mutually unbiased, $|\langle b_k|b'_l\rangle|^2=1/d$~\cite{Tavakoli2015,Aguilar2018}, and we assume this except where stated otherwise. With $M_{ij}=\Pi_i+\Pi'_j$ and $s_{ij}=\tfrac12\Tr(\rho_{ij}M_{ij})$, the success probability is $P_{\rm PM}=d^{-2}\sum_{ij}s_{ij}$. Sending one of the two dits gives the classical value $P_C(d)=\tfrac12(1+1/d)$, and no classical $d$-level message does better, even with shared randomness~\cite{Tavakoli2015,Ambainis2024}. A code has a quantum advantage if $P_{\rm PM}>P_C(d)$. For mutually unbiased bases the largest eigenvalue of $M_{ij}$ is $1+1/\sqrt d$, which gives the quantum value $P_Q(d)=\tfrac12(1+1/\sqrt d)$~\cite{Tavakoli2015,Aguilar2018}. For $d=2$ we take $B=\sigma_z$ and $B'=\sigma_x$ without loss of generality: every pair of qubit MUBs is obtained from this one by a unitary, which leaves the success probability and the irrealism unchanged when applied to states and measurements together. Up to such a rotation, this is the $2\to1$ code that underlies semi-device-independent cryptography~\cite{Pawlowski2011} and randomness expansion~\cite{Li2011}, and it has been mapped onto an interferometric wave--particle test whose witness has been measured with photons~\cite{Raj2026}.

\emph{Entanglement-assisted codes.} In the EA code~\cite{Pawlowski2010} Alice and Bob share $\rho_{\cA\cB}$, with $\cA$ a quantum system of arbitrary finite dimension and $\cB$ a qubit; a quantum description of the shared resource is what makes Eq.~\eqref{eq:irr} applicable to Bob's observables. Alice measures $A_z$, with $z=x_0\oplus x_1$, $A_0\equiv A$ and $A_1\equiv A'$, obtains $a\in\{0,1\}$, and sends $m=x_0\oplus a$. Bob measures $B=\sigma_z$ ($y=0$) or $B'=\sigma_x$ ($y=1$), which, as in the PM code, entails no loss of generality, obtains $b$, and outputs $m\oplus b$, which is correct if and only if $b=a$ for $y=0$ and $b=a\oplus z$ for $y=1$. Assigning the eigenvalues $(-1)^a$ and $(-1)^b$ to the outcomes, $P_{\rm EA}=\tfrac12+\cS/8$, with $\cS=\langle A\otimes(B+B')\rangle+\langle A'\otimes(B-B')\rangle$ the CHSH value~\cite{CHSH1969}; local models give $\cS\le2$, i.e., $P_{\rm EA}\le3/4$, and quantum mechanics allows $\cS\le2\sqrt2$~\cite{Tsirelson1980}, attained in Ref.~\cite{Pawlowski2010}. Bob's measurement depends only on $y$, which makes $B$ and $B'$ the decoding observables, and Alice's only on $z$, so that, conditioned on her setting and outcome, Bob's conditional states form a steered code, an ensemble with the same decoders to which the PM analysis applies (Sec.~\ref{sec:EA}). Protocols in which Alice's setting depends on both bits~\cite{Kanjilal2023} fall outside this class, and in higher dimensions the two architectures can differ~\cite{Tavakoli2016}.

\subsection{Irrealism and the information--reality exclusion}\label{sec:exclusion}

With $\Phi_B$ the map of Eq.~\eqref{eq:criterion}, acting on $\rho_{\cA\cB}$ or on Bob's reduced state, the irrealism splits into a local and a correlation part~\cite{Bilobran2015},
\begin{equation}
\Irr_B(\rho_{\cA\cB})=\Irr_B(\rho_\cB)+\Big[I(\cA{:}\cB)_{\rho}-I(\cA{:}\cB)_{\Phi_B(\rho)}\Big],
\label{eq:BA}
\end{equation}
since $\Tr_\cB\Phi_B(\rho_{\cA\cB})=\rho_\cA$. The first term is the relative entropy of coherence of Bob's reduced state in the eigenbasis of $B$~\cite{Baumgratz2014,Streltsov2017}; the second is the mutual information destroyed by an unrevealed measurement of $B$, a discord-like quantity attached to $B$ rather than optimized over measurements. The quantum discord with measurements on $\cB$~\cite{OllivierZurek2001,HendersonVedral2001} does not exceed the second term, in which the measurement is fixed to $B$, so that $\Irr_B(\rho_{\cA\cB})$ is bounded from below by the local coherence plus the discord. An observable can thus lack a definite value even when Bob's local state is diagonal in its eigenbasis, if its value is not fixed relative to Alice's system; Eq.~\eqref{eq:BA} holds for any system in the role of $\cA$, including a classical register.

A measurement on Alice's side, whose outcome is not revealed, can also change the reality of $B$, although it leaves Bob's reduced state unchanged. For a nondegenerate observable $A$ of $\cA$, with $\Phi_A$ defined as in Eq.~\eqref{eq:criterion} with the roles of $\cA$ and $\cB$ exchanged, the realism-based nonlocality of the context $(A,B)$, a pair of observables, one for each party~\cite{Bilobran2015}, is
\begin{equation}
\eta_{B|A}(\rho_{\cA\cB})=\Irr_B(\rho_{\cA\cB})-\Irr_B\big(\Phi_A(\rho_{\cA\cB})\big)\ \ge\ 0,
\label{eq:rbn}
\end{equation}
nonnegative by data processing, since $\Phi_A$ commutes with $\Phi_B$. Because $\Tr_\cA\Phi_A(\rho_{\cA\cB})=\rho_\cB$, Alice's measurement affects only the correlation term of Eq.~\eqref{eq:BA}. Expanding both irrealisms, $\eta_{B|A}=S(\Phi_A(\rho_{\cA\cB}))+S(\Phi_B(\rho_{\cA\cB}))-S(\rho_{\cA\cB})-S(\Phi_A\Phi_B(\rho_{\cA\cB}))$, which is symmetric, $\eta_{B|A}=\eta_{A|B}$: the reality that an unrevealed measurement of $A$ confers on $B$ equals the reality that one of $B$ confers on $A$. Its maximum over the nondegenerate observables of both parties, $\eta_{\max}(\rho_{\cA\cB})=\max_{A,B}\eta_{B|A}(\rho_{\cA\cB})$, is the realism-based nonlocality of the state~\cite{Gomes2018}, which can be positive even for separable states~\cite{Bilobran2015}.

Following Ref.~\cite{Dieguez2018}, the information in the preparation is $\Info(\rho_{\cA\cB})=\log_2(d_\cA d_\cB)-S(\rho_{\cA\cB})$, where $d_\cA$ and $d_\cB$ are the dimensions of the two subsystems. Calling $\Info_B(\rho_{\cA\cB})=\log_2(d_\cA d_\cB)-S(\Phi_B(\rho_{\cA\cB}))$ the information about $B$, the part of $\Info(\rho_{\cA\cB})$ that survives an unrevealed measurement of $B$, one has
\begin{equation}
\Info(\rho_{\cA\cB})=\Info_B(\rho_{\cA\cB})+\Irr_B(\rho_{\cA\cB}).
\label{eq:complementarity}
\end{equation}
Since monitoring of $B$ leaves $\Info_B$ unchanged, the reality it confers on $B$ is paid for by an equal loss of information~\cite{Dieguez2018}.

For two mutually unbiased observables $B$ and $B'$ of Bob's system, Eq.~\eqref{eq:complementarity} becomes an exclusion between them. Since $\Phi_{B'}(\Phi_B(\rho_{\cA\cB}))=\rho_\cA\otimes\id/d_\cB$, which uses $|\langle b_k|b'_l\rangle|^2=1/d_\cB$, applying $\Phi_{B'}$ to both arguments of $\Irr_B(\rho_{\cA\cB})=D(\rho_{\cA\cB}\Vert\Phi_B(\rho_{\cA\cB}))$, which cannot increase it~\cite{Lindblad1975,Uhlmann1977}, gives $\Irr_B(\rho_{\cA\cB})\ge D(\Phi_{B'}(\rho_{\cA\cB})\Vert\rho_\cA\otimes\id/d_\cB)$, that is,
\begin{equation}
\Irr_B(\rho_{\cA\cB})\ \ge\ \Info_{B'}(\rho_\cB)+I(\cA{:}\cB)_{\Phi_{B'}(\rho)}.
\label{eq:exclusion}
\end{equation}
The first term, $\Info_{B'}(\rho_\cB)=\log_2d_\cB-H_{B'}$, with $H_{B'}=-\sum_lp'_l\log_2p'_l$, $p'_l=\Tr(\rho_\cB\Pi'_l)$, the Shannon entropy of the outcome of $B'$, is the information about $B'$ in Bob's reduced state; the second is the part of the correlations between Alice and Bob that survives an unrevealed measurement of $B'$, that is, the correlation of $B'$ with Alice's system. If $B$ is real, $B'$ carries no information. For a single system, Eq.~\eqref{eq:exclusion} reads $H_B+H_{B'}\ge\log_2d_\cB+S(\rho_\cB)$, with $H_B$ the Shannon entropy of the outcome of $B$, defined likewise, the uncertainty relation of Refs.~\cite{Berta2010,Korzekwa2014} without quantum memory, which for pure states reduces to the Maassen--Uffink relation~\cite{Maassen1988,Coles2017}. In general, with $E$ a system that purifies $\rho_{\cA\cB}$, $\Irr_B(\rho_{\cA\cB})=H(B|E)$ (Sec.~\ref{sec:gibbs}), where $H(X|Y)=H(XY)-H(Y)$ is the conditional entropy, $H$ is the entropy in bits of the classical outcomes and quantum systems involved, and $B$ in such expressions stands for the outcome of $B$ recorded in a classical register; and the right-hand side of Eq.~\eqref{eq:exclusion} equals $\log_2d_\cB-H(B'|\cA)$, with $H(B'|\cA)$ the conditional entropy of the outcome of $B'$ given Alice's system. Equation~\eqref{eq:exclusion} is therefore the tripartite uncertainty relation with quantum memory, $H(B|E)+H(B'|\cA)\ge\log_2d_\cB$ for mutually unbiased observables~\cite{RenesBoileau2009,Berta2010}, and the derivation above is the data-processing proof of Ref.~\cite{Coles2012PRL}. Read in terms of irrealism, it states that information about one observable, in Bob's system or in its correlations with Alice's, can exist only to the extent that the other is not real; we refer to it as the information--reality exclusion. If Alice performs a measurement $\{E_a\}$ on her system, which cannot increase the mutual information, Eq.~\eqref{eq:exclusion} gives a bound in terms of observed statistics,
\begin{equation}
\Irr_B(\rho_{\cA\cB})\ \ge\ \Info_{B'}(\rho_\cB)+I(a{:}b'),
\label{eq:exclusion2}
\end{equation}
with $I(a{:}b')$ the mutual information of $p(a,b')=\Tr[\rho_{\cA\cB}(E_a\otimes\Pi'_{b'})]$.

With Alice's system as a quantum memory, the sum of the two irrealisms obeys
\begin{equation}
\Irr_B(\rho_{\cA\cB})+\Irr_{B'}(\rho_{\cA\cB})\ \ge\ \log_2d_\cB-S(\cB|\cA)_\rho,
\label{eq:memory}
\end{equation}
with $S(\cB|\cA)_\rho=S(\rho_{\cA\cB})-S(\rho_\cA)$, which is the bipartite uncertainty relation of Ref.~\cite{Berta2010} written in irrealism form~\cite{Freire2019,Dieguez2022,Rudnicki2018}; its right-hand side equals $S(\rho_{\cA\cB}\Vert\rho_\cA\otimes\id/d_\cB)$. Equation~\eqref{eq:exclusion2} bounds each irrealism through the classical correlations that Alice can reveal, by at most $\log_2d_\cB$; Eq.~\eqref{eq:memory} bounds only the sum, but through the full quantum correlations, and reaches $2\log_2d_\cB$ for a maximally entangled state.

\section{The irrealism frontier}\label{sec:cost}

\subsection{The question}\label{sec:question}

The relations of Sec.~\ref{sec:exclusion} bound the irrealism of the decoding observables without determining it. We now ask how much irrealism the decoding observables must have for a code to reach a given success probability. In both architectures Bob decodes a system that is correlated with Alice's side: with her input register in PM codes, and with her half of the shared resource in EA codes. We describe both by a state $\rho_{\cA\cB}$ of Alice's side and Bob's system, taken before any measurement, and measure the irrealism of the decoding observables by their total irrealism,
\begin{equation}
\cC=\Irr_B(\rho_{\cA\cB})+\Irr_{B'}(\rho_{\cA\cB}),
\label{eq:total}
\end{equation}
which we denote $\cC_{\rm PM}$ and $\cC_{\rm EA}$ in the two architectures. Bob's decoding observables are rank-one projective, as the definition of irrealism requires; Alice is free to choose how she encodes her input. For given decoders, the least total irrealism among the codes whose success probability $P_{\rm succ}$, that is, $P_{\rm PM}$ or $P_{\rm EA}$, reaches a target value $P$ is
\begin{equation}
\cC_{\min}(P\,|\,B,B')=\inf\big\{\cC:\ P_{\rm succ}\ge P\big\},
\label{eq:question}
\end{equation}
where the infimum runs over Alice's strategies: her encodings in PM codes, the shared state and her measurements in EA codes. We bound this sum, which is what the relations of Sec.~\ref{sec:exclusion} constrain, and each irrealism separately.

The decoders are part of the question, because irrealism, like coherence, is defined relative to given observables. We take those of the optimal code: the quantum optimum is attained only by rank-one projective decoders in mutually unbiased bases, in PM codes~\cite{Farkas2019} and, through the rigidity of the CHSH inequality~\cite{Supic2020}, in EA codes with a qubit on Bob's side. We call $\cC^{(d)}_{\min}(P)$ and $\cC^{\rm EA}_{\min}(P)$, Eq.~\eqref{eq:question} with MUB decoders, the irrealism frontiers of PM and EA codes. Section~\ref{sec:decoders} determines what this choice costs, and what remains when the decoders are optimized as well. 

\subsection{Prepare-and-measure codes}\label{sec:PM}

In PM codes, Alice's side is her input register $X$, and the state of Eq.~\eqref{eq:total} is
\begin{equation}
\omega_{X\cB}=\frac1{d^2}\sum_{ij}\proj{ij}\otimes\rho_{ij}.
\label{eq:cq}
\end{equation}
Since $\Phi_B$ acts only on Bob's system, $S(\omega_{X\cB})$ and $S(\Phi_B(\omega_{X\cB}))$ both equal the entropy $\log_2d^2$ of the input plus the average entropy of the conditional states, $\rho_{ij}$ or $\Phi_B(\rho_{ij})$. The irrealism of this classical--quantum state is therefore the average irrealism of its conditional states, and the total irrealism is the average over messages, $\cC_{\rm PM}=d^{-2}\sum_{ij}[\Irr_B(\rho_{ij})+\Irr_{B'}(\rho_{ij})]$, which is also the average sum of the relative entropies of coherence of the transmitted states in the two decoding bases.

We first treat each decoding observable separately, which already locates the threshold at the classical value, and then the total irrealism, for which the frontier takes closed form. A decoding observable can be an element of reality only up to the classical value, a short lemma that is the counterpart, for the transmitted states, of the need for incompatible decoders~\cite{Carmeli2020}, and is closely related to the use of QRACs as witnesses of coherence~\cite{SarkarDatta2025}. Suppose that, for arbitrary rank-one projective decoders, $B$ is an element of reality for every transmitted state, so that $\rho_{ij}=\sum_kp(k|ij)\Pi_k$. Bob's answers then depend on the message only through the label $k$: Alice could send $k$ as a classical $d$-level message, drawn with probability $p(k|ij)$, and Bob could simulate his outcome of $B'$ with probabilities $\Tr(\Pi_k\Pi'_l)$. The code is therefore classical and $P_{\rm PM}\le P_C(d)$; the same holds with the roles of $B$ and $B'$ exchanged.

The exclusion also bounds the irrealism that the advantage requires of each observable. Let $p_{ij}=\Tr(\rho_{ij}\Pi_i)$ and $q_{ij}=\Tr(\rho_{ij}\Pi'_j)$, so that $s_{ij}=(p_{ij}+q_{ij})/2$. A $d$-outcome distribution with one entry equal to $q$ has entropy at most $h_2(q)+(1-q)\log_2(d-1)$, with $h_2(q)=-q\log_2q-(1-q)\log_2(1-q)$ the binary entropy, so $\Info_{B'}(\rho_{ij})\ge K_d(q_{ij})$ and $\Info_B(\rho_{ij})\ge K_d(p_{ij})$, with
\begin{equation}
K_d(q)=\log_2d-h_2(q)-(1-q)\log_2(d-1).
\label{eq:Kdef}
\end{equation}
Applying Eq.~\eqref{eq:exclusion} to each transmitted state, for which Alice's side is trivial, gives $\Irr_B(\rho_{ij})\ge \Info_{B'}(\rho_{ij})\ge K_d(q_{ij})$; averaged over messages, this is Eq.~\eqref{eq:exclusion} for the state~\eqref{eq:cq}, with the correlation with Alice's input included. Since $q_{ij}\ge2s_{ij}-1$, and $K_d$ vanishes at $q=1/d$ and increases for $q\ge1/d$, $\Irr_B(\rho_{ij})\ge k_d(s_{ij})$, where $k_d(s)=K_d(2s-1)$ for $s\ge P_C(d)$ and $k_d(s)=0$ otherwise; $k_d$ is convex and nondecreasing, because $K_d$ is convex with zero slope at $q=1/d$. Jensen's inequality over messages then gives, for $P_{\rm PM}\ge P_C(d)$, which is equivalent to $2P_{\rm PM}-1\ge1/d$,
\begin{equation}
\frac1{d^2}\sum_{ij}\Irr_B(\rho_{ij})\ \ge\ K_d(2P_{\rm PM}-1),
\label{eq:Kd}
\end{equation}
where $K_d$ vanishes with zero slope at $q=1/d$; the same bound holds for $B'$ (Appendix~\ref{app:seqone}). Below $P_C(d)$ the bound is zero, as it must be, since a decoding observable can then be real. Since $K_d(q)>0$ for every $q>1/d$, the bound~\eqref{eq:Kd} is positive exactly in the advantage region $P_{\rm PM}>P_C(d)$.

This bound is not tight, but the exact minimum is accessible in every dimension. For one message, a unitary diagonal in the eigenbasis of $B$ maps $\ket{b'_j}$ to $d^{-1/2}\sum_k\ket{b_k}$ without changing $\Irr_B$ or $\Pi_i$, and the permutations of the eigenvectors $\ket{b_k}$ with $k\neq i$ then leave the problem invariant. By the convexity of $\Irr_B$ the minimum is attained by an invariant state, whose outcome distribution for $B$ has entropy $f_d(p)=h_2(p)+(1-p)\log_2(d-1)$, with $p=\Tr\rho\Pi_i$. Since $\Irr_B=S(\Phi_B(\rho))-S(\rho)$, what remains is to maximize $S(\rho)$ at fixed $\Tr\rho\Pi_i$ and $\Tr\rho\Pi'_j$, which the Gibbs states $\rho_{\lambda\mu}=e^{\lambda\Pi_i+\mu\Pi'_j}/\Tr e^{\lambda\Pi_i+\mu\Pi'_j}$ do (Appendix~\ref{app:seqone}). For $1/d\le P\le P_Q$, the minimum $\Irr^{\min}_{\rm PM}(P)$ of the average irrealism $d^{-2}\sum_{ij}\Irr_B(\rho_{ij})$ of one decoding observable over codes with $P_{\rm PM}\ge P$ is therefore
\begin{equation}
\begin{aligned}
\Irr^{\min}_{\rm PM}(P)=\inf_{\lambda,\mu}\Big\{&f_d\big(\Tr\rho_{\lambda\mu}\Pi_i\big)-S(\rho_{\lambda\mu}):\\&\tfrac12\Tr\big[\rho_{\lambda\mu}(\Pi_i+\Pi'_j)\big]=P\Big\},
\end{aligned}
\label{eq:seqoned}
\end{equation}
a one-parameter minimization. Since each pair $(p,q)$ with $q=\Tr\rho_{\lambda\mu}\Pi'_j$ in the interior of the attainable set fixes one Gibbs state, it reads
\begin{equation}
\Irr^{\min}_{\rm PM}(P)=\min_p\big[f_d(p)-S_{\max}(p,2P-p)\big],
\label{eq:seqonep}
\end{equation}
where $S_{\max}(p,q)=S(\rho_{\lambda\mu})$ is the largest entropy at fixed $p$ and $q$, and $p$ runs over the values for which $(p,2P-p)$ is attainable, an interval that contains $P$ and shrinks to $\{P_Q\}$ as $P\to P_Q$. The same holds for $B'$. It vanishes if and only if $P\le P_C(d)$, is strictly positive beyond, and reaches $f_d(P_Q)=\cC_Q(d)/2$ at $P_Q$: in every dimension, a decoding observable can remain an element of reality exactly up to the classical value. The minimizing states are strongly asymmetric: they keep the decoding probability $\Tr\rho_{ij}\Pi_i$ of $B$ close to one, so that $B$ remains nearly real, and draw the additional success from $B'$; the two decoding probabilities meet only at $P_Q$. For $d=2$, Eq.~\eqref{eq:seqoned} takes the explicit form
\begin{equation}
\begin{aligned}
\Irr^{\min}_{\rm PM}(P)=\min_{|\epsilon|\le\omega}\Big[&h_2\Big(\frac{2+\upsilon+2\epsilon}{4}\Big)\\&-h_2\Big(\frac{1+\sqrt{\upsilon^2/2+2\epsilon^2}}{2}\Big)\Big],
\end{aligned}
\label{eq:seqone}
\end{equation}
with $\upsilon=2(2P-1)$ and $\omega=(1/2-\upsilon^2/4)^{1/2}$. At $P_Q$, where $\omega=0$, the minimum equals $h_2(\tfrac12+\tfrac1{2\sqrt2})\simeq0.601$. For $P=0.8$, for instance, Eq.~\eqref{eq:Kd} gives $0.029$ against the exact value $0.094$.

We now turn to the total irrealism. For both decoding observables together, Eq.~\eqref{eq:exclusion} becomes a lower bound on it: applied to both observables, it gives $\Irr_B(\rho_{ij})+\Irr_{B'}(\rho_{ij})\ge K_d(p_{ij})+K_d(q_{ij})\ge2K_d(s_{ij})$, by the convexity of $K_d$, and Jensen's inequality over messages gives
\begin{equation}
\cC_{\rm PM}\ \ge\ 2K_d(P_{\rm PM}).
\label{eq:exbound}
\end{equation}
The right-hand side is a lower bound on the information about the two decoding observables that a code with success probability $P_{\rm PM}$ must carry, and Eq.~\eqref{eq:exclusion} charges it to the irrealism of the other observable. It vanishes only at $P_{\rm PM}=1/d$; for $d=2$, $K_2(q)=1-h_2(q)$. Section~\ref{sec:gibbs} shows by how much the exact minimum exceeds it.

\subsubsection{Exact frontier}\label{sec:gibbs}

Unlike the irrealism of each decoding observable, which can vanish up to the classical value, the total irrealism is positive, by the exclusion bound~\eqref{eq:exbound}, at every success probability above the guessing value $1/d$; it does not switch on at the classical value. Computing its exact minimum requires a variational principle.

The principle is an identity for the total irrealism $F=\Irr_B+\Irr_{B'}$ of a single state. Suppose that a full-rank state $\rho$ satisfies the closure property
\begin{equation}
2\log\rho-\log\Phi_B(\rho)-\log\Phi_{B'}(\rho)=\kappa\id+\zeta M
\label{eq:closure}
\end{equation}
for a Hermitian $M$ and real $\kappa,\zeta$. Then every state $\sigma$ with $\Tr(\sigma M)=\Tr(\rho M)$ obeys
\begin{equation}
F(\sigma)-F(\rho)=\sum_{\Phi=\Phi_B,\Phi_{B'}}\Big[D(\sigma\Vert\rho)-D\big(\Phi(\sigma)\Vert\Phi(\rho)\big)\Big]\ \ge\ 0 .
\label{eq:excess}
\end{equation}
The excess irrealism of any competitor is the distinguishability from the candidate that the unrevealed measurements erase, and it is nonnegative by data processing. If equality holds, Petz's characterization of equality in the data-processing inequality~\cite{Petz1986,JencovaPetz2006,Hiai2011} forces $\rho^{-1/2}\sigma\rho^{-1/2}$ to commute with the projectors of both $B$ and $B'$ (Appendix~\ref{app:lemma}). The closure property thus turns a variational problem into a check: the unrevealed measurements must map the candidate optimum into the span of the identity and the constrained observable. The nontrivial step in a given problem is to find a closed point in closed form.

\begin{figure*}[t]
\centering
\includegraphics[width=\textwidth]{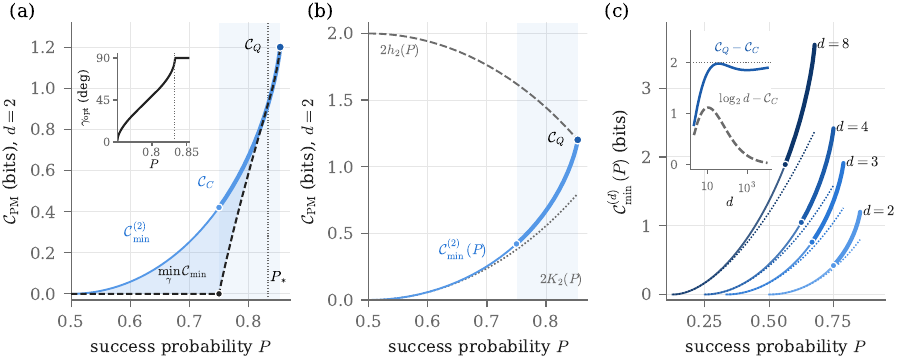}
\caption{Minimum total irrealism of PM codes. Thin lines: $P\le P_C(d)$; thick lines: codes with a quantum advantage; dots mark the classical value $P_C(d)$; in (a) and (b) the light band is the quantum-advantage region. (a) Qubit codes with the decoders fixed to MUBs, $\cC^{(2)}_{\min}$ [Eq.~\eqref{eq:qubit}], and with the decoders also optimized, $\min_\gamma\cC_{\min}(P\,|\,\gamma)$ (dashed), obtained by minimizing the exact expression~\eqref{eq:anydec} numerically over the angle $\gamma$ between the Bloch axes of the decoders. The latter vanishes up to $P=3/4$, where the MUB frontier is still positive, lies strictly below the MUB frontier also for $3/4<P<P_*\simeq0.833$, and coincides with it numerically above $P_*$; the shaded area between the two is the additional irrealism required by mutually unbiased decoders. Inset: the optimal angle $\gamma_{\rm opt}$, which rises from $0$ at $P=3/4$ to $90^\circ$ at $P_*$. (b) Qubit codes: the frontier, Eq.~\eqref{eq:qubit}, lies between the exclusion bound $2K_2(P)$ (dotted) and the upper bound $2h_2(P)$ obeyed by every code (dashed), Eq.~\eqref{eq:window}; the frontier and $2h_2(P)$ meet at the quantum optimum, where every code has total irrealism $\cC_Q(2)\simeq1.20$ bits. (c) Exact minimum total irrealism for MUBs in dimensions $d=2,3,4,8$ (light to dark), Eq.~\eqref{eq:frontier}; dotted curves: exclusion bound $2K_d(P)$, Eq.~\eqref{eq:exbound}. Inset (bits; logarithmic $d$ axis): the gap $\cC_Q-\cC_C$ between the quantum and classical endpoints, which tends to $2$ bits, and the difference $\log_2d-\cC_C$ between the classical strategy and the optimal code at the same performance, which tends to $0$ [Eq.~\eqref{eq:ICas}].}
\label{fig:dim}
\end{figure*}

For MUBs in dimension $d\ge2$ the natural candidate is the Gibbs state of the decoding operator $M_{ij}=\Pi_i+\Pi'_j$,
\begin{equation}
\rho_{ij}(\theta)=\frac{e^{\theta(\Pi_i+\Pi'_j)}}{Z_d(\theta)},\qquad Z_d(\theta)=d-2+2e^{\theta}\cosh\frac{\theta}{\sqrt d},
\label{eq:gibbs}
\end{equation}
with $\theta\in\mathbb R$; when the message is fixed we write $\rho_\theta$. The operator $M_{ij}$ has eigenvalues $1\pm d^{-1/2}$ and a $(d-2)$-dimensional kernel, which fixes $Z_d$. A reflection that exchanges $\ket{b_i}$ and $\ket{b'_j}$, up to a phase, leaves $M_{ij}$ invariant, so the two decoding probabilities are equal, $\Tr[\rho_{ij}(\theta)\Pi_i]=\Tr[\rho_{ij}(\theta)\Pi'_j]=P(\theta)$, with
\begin{equation}
P(\theta)=\frac{e^{\theta}\big[\cosh(\theta/\sqrt d)+d^{-1/2}\sinh(\theta/\sqrt d)\big]}{Z_d(\theta)}.
\label{eq:Ptheta}
\end{equation}
Since $dP/d\theta=\tfrac12\mathrm{Var}_{\rho_\theta}(M_{ij})>0$, this is a strictly increasing bijection from $\mathbb R$ onto $(0,P_Q)$ for $d>2$ and onto $(\tfrac12-\tfrac1{2\sqrt2},P_Q)$ for $d=2$, with $P(0)=1/d$; only $\theta\ge0$, i.e., $P\ge1/d$, enters the frontier.

What makes the Gibbs state optimal is the closure property. For every $k\neq i$, the diagonal element $\bra{b_k}\rho_{ij}(\theta)\ket{b_k}$ is the same, because $e^{\theta M_{ij}}$ is a quadratic polynomial in $M_{ij}$ (linear for $d=2$) and $\bra{b_k}M_{ij}\ket{b_k}=\bra{b_k}M_{ij}^2\ket{b_k}=1/d$ by unbiasedness. Hence
\begin{equation}
\Phi_B(\rho_{ij})=P(\theta)\Pi_i+\frac{1-P(\theta)}{d-1}(\id-\Pi_i),
\label{eq:deph}
\end{equation}
and similarly for $B'$. The left-hand side of Eq.~\eqref{eq:closure} is therefore $[2\theta-\ln\tfrac{(d-1)P}{1-P}]M_{ij}$ plus a multiple of $\id$, and Eq.~\eqref{eq:excess} with $M=M_{ij}$ shows that $\rho_{ij}(\theta)$ minimizes $\Irr_B+\Irr_{B'}$ among \emph{all} states with the same $s_{ij}$. Only the sum of the two decoding probabilities is constrained, so no symmetrization between the bases is needed. Equality forces $\rho^{-1/2}\sigma\rho^{-1/2}$ to be diagonal in both bases, hence proportional to $\id$ because $\langle b_k|b'_l\rangle\neq0$ for all $k,l$, and the minimizer is unique. The single-message minimum is a convex function of $s_{ij}$, the same for every message, so Jensen's inequality extends the result to the whole code (Appendix~\ref{app:gibbs}).

\emph{Result (PM frontier).} For PM $2_d\to1$ codes decoded in MUBs and $1/d\le P\le P_Q(d)$, the minimum of $\cC_{\rm PM}$ over all codes with $P_{\rm PM}\ge P$ is
\begin{equation}
\begin{aligned}
\cC^{(d)}_{\min}(P)=2\Big[&h_2(P)+(1-P)\log_2(d-1)\\&-\log_2Z_d(\theta)+\frac{2\theta P}{\ln2}\Big],
\end{aligned}
\label{eq:frontier}
\end{equation}
with $\theta\ge0$ the solution of $P(\theta)=P$. For $1/d<P<P_Q$ the optimal code is unique: $\rho_{ij}=\rho_{ij}(\theta)$ for all $(i,j)$, with the same $\theta$. At $P=P_Q$ the optimal code is the standard pure code, and $\cC_Q(d)\equiv \cC^{(d)}_{\min}(P_Q)=2[h_2(P_Q)+(1-P_Q)\log_2(d-1)]$. The frontier depends on the bases only through $|\langle b_k|b'_l\rangle|^2=1/d$, so it is the same for every pair of MUBs, including inequivalent pairs for $d\ge4$. A covariance inequality shows that the frontier is strictly convex and increasing, with slope
\begin{equation}
\frac{d\cC^{(d)}_{\min}}{dP}=\frac{2}{\ln2}\Big[2\theta-\ln\frac{(d-1)P}{1-P}\Big],
\label{eq:slope}
\end{equation}
which vanishes at $P=1/d$.

The frontier has an upper partner. For each message, $\Irr_B+\Irr_{B'}=H_B+H_{B'}-2S(\rho_{ij})$, and a $d$-outcome distribution whose intended outcome has probability $p$ has entropy at most $f_d(p)$, a concave function of $p$. Jensen's inequality over the $2d^2$ decoding probabilities then gives, for every code (Appendix~\ref{app:gibbs}),
\begin{equation}
\cC^{(d)}_{\min}(P_{\rm PM})\ \le\ \cC_{\rm PM}\ \le\ 2f_d(P_{\rm PM})=\cC^{(d)}_{\min}(P_{\rm PM})+2S(\rho_\theta),
\label{eq:window}
\end{equation}
where the equality is Eq.~\eqref{eq:frontier} rewritten, with $\rho_\theta$ the Gibbs state at $P(\theta)=P_{\rm PM}$. The Gibbs code saturates the marginal entropies, $H_B=H_{B'}=f_d(P)$ by Eq.~\eqref{eq:deph}, while having the largest von Neumann entropy compatible with the mean of $M_{ij}$. The upper bound is attained for qubits at every $P\in[\tfrac12,P_Q]$, by pure states with equal decoding probabilities, and in every dimension at $P_Q$. Whether it is attained for $d>2$ and $1/d<P<P_Q$ we leave open; for the Fourier pair of bases with $3\le d\le7$, a numerical search finds, at every $P$ tested, a pure state with the required outcome distributions, which the Weyl--Heisenberg displacements carry to the other messages. The width $2S(\rho_\theta)$ of the window decreases monotonically to zero as $P\to P_Q$ [Fig.~\ref{fig:dim}(b)]. Far from the optimum the success probability leaves the irrealism largely undetermined; at the optimum it fixes it, $\cC_{\rm PM}=\cC_Q(d)$ for every code, and with it the irrealism of each decoding observable, $\Irr_B=\Irr_{B'}=f_d(P_Q)$, because for mutually unbiased decoders $P_Q(d)$ is reached only if every $\rho_{ij}$ is the eigenvector of $M_{ij}$ with the largest eigenvalue, which is nondegenerate.

The exact minimum exceeds the exclusion bound~\eqref{eq:exbound} [Fig.~\ref{fig:dim}(c)]. At the classical value the two are $0.421$ and $0.377$ bits for $d=2$, and $0.761$ and $0.667$ for $d=3$; at the quantum optimum they are $1.202$ and $0.798$ for $d=2$, and $1.911$ and $1.259$ for $d=3$. Their difference is of fourth order in $P-1/d$, where little information is needed, and the exact minimum lies strictly above the bound for every $P>1/d$ and every $d$ (Appendix~\ref{app:gibbs}). For large $d$ the difference tends to two bits at the classical value and to four bits at the quantum optimum (see below). The exclusion therefore accounts for the irrealism needed near the guessing value, but not for the excess required at higher performance.

The optimal code is thus a Gibbs code: each $\rho_{ij}(\theta)$ is also the maximum-entropy state with the given mean of $M_{ij}$. It is the closure property, however, that makes this state optimal for irrealism; for generic bases the two notions need not coincide (Sec.~\ref{sec:conclusion}). Closure says that the dephased optimum, Eq.~\eqref{eq:deph}, is determined by the decoding probability alone: the unrevealed measurements erase every feature of the state except the probability that Bob decodes correctly. Mutual unbiasedness enters the proof through the spectrum of $M_{ij}$, through the equal diagonal elements $\bra{b_k}M_{ij}\ket{b_k}=\bra{b_k}M_{ij}^2\ket{b_k}=1/d$ that give closure, and through the nonvanishing overlaps that give uniqueness; the exclusion~\eqref{eq:exclusion} is not used. Three consequences deserve comment. First, the total irrealism vanishes only if every $\rho_{ij}=\id/d$, since a state for which both $B$ and $B'$ are elements of reality is proportional to $\id$; accordingly, $\cC^{(d)}_{\min}(P)=0$ only at $P=1/d$, and the frontier is smooth through $P_C(d)$ [Fig.~\ref{fig:dim}(c)]. Second, the optimal codes are mixed except at $P_Q$. Uniqueness excludes, in particular, the isotropic family $(1-\lambda)\id/d+\lambda\proj{\psi_{ij}}$, with $\ket{\psi_{ij}}$ the standard optimal state, for every $d>2$ and every $P\in(1/d,P_Q)$, since its spectrum is $(d-1)$-fold degenerate whereas that of $\rho_{ij}(\theta)$ is not. Third, the classical strategy of sending $i$ has total irrealism exactly $\log_2d$, with $\Irr_B=0$ and $\Irr_{B'}=\log_2 d$, strictly more than $\cC^{(d)}_{\min}(P_C)$.

For qubits the frontier follows as a corollary of Eq.~\eqref{eq:frontier}, which can also be checked by a direct minimization over Bloch vectors. For $d=2$, Eq.~\eqref{eq:Ptheta} becomes $P=\tfrac12[1+\tanh(\theta/\sqrt2)/\sqrt2]$, the optimal states are $\rho=\tfrac12(\id+r\,\hat n\cdot\bm\sigma)$ with $\hat n=(\hat x+\hat z)/\sqrt2$ and $r=\sqrt2(2P-1)$, and the frontier takes the closed form
\begin{equation}
\cC^{(2)}_{\min}(P)=2\Big[h_2(P)-h_2\Big(\frac12+\frac{2P-1}{\sqrt2}\Big)\Big].
\label{eq:qubit}
\end{equation}
For qubits the isotropic and Gibbs families coincide.

The total minimum is not twice the minimum for one decoding observable, Eq.~\eqref{eq:seqone}. The Gibbs code has $r_z=r_x$, i.e., $\epsilon=0$ in Eq.~\eqref{eq:seqone}, and $\Irr_B=\Irr_{B'}$, so $\cC^{(2)}_{\min}(P)$ is twice the bracket of Eq.~\eqref{eq:seqone} at $\epsilon=0$. For $\tfrac12<P<P_Q$ the point $\epsilon=0$ lies inside the interval $|\epsilon|\le\omega$, and the derivative of the bracket there is $\tfrac12\log_2[(2-\upsilon)/(2+\upsilon)]<0$ (Appendix~\ref{app:seqone}). Hence
\begin{equation}
\cC^{(2)}_{\min}(P)\ >\ 2\,\Irr^{\min}_{\rm PM}(P),\qquad \tfrac12<P<P_Q.
\label{eq:strict}
\end{equation}
The two minima are reached by different states. For one message at fixed $P$, the attainable pairs $(\Irr_B,\Irr_{B'})$ form a region that reaches the axis $\Irr_B=0$ only for $P\le P_C$ and never contains the origin, since the only state for which both decoding observables are elements of reality is $\id/d$. The minimum for one observable is the point of this region with the smallest $\Irr_B$, reached by strongly asymmetric states in which the other observable carries most of the irrealism. The minimum total is the point at which a line along which the sum of the two irrealisms is constant first touches the region; it lies on the diagonal $\Irr_B=\Irr_{B'}$ and is reached by the Gibbs state (Appendix~\ref{app:seqone}). At $P=0.8$, for instance, the state that minimizes $\Irr_B$ has $\Irr_B=0.094$ and $\Irr_{B'}=0.92$ bits, whereas the Gibbs state has $0.335$ bits for each.

Because $\cC^{(d)}_{\min}$ is continuous and strictly increasing on $[1/d,P_Q(d)]$ (at $P_Q$, $S(\rho_\theta)\to0$ as $\theta\to\infty$), every PM $2_d\to1$ code with MUBs and a quantum advantage satisfies
\begin{equation}
\cC_{\rm PM}\ >\ \cC_C(d)\equiv \cC^{(d)}_{\min}\big(P_C(d)\big),
\label{eq:classval}
\end{equation}
and $\cC_C(d)$, the minimum total irrealism at the classical value, is the infimum over such codes, approached by the Gibbs codes. Read in reverse, a code with $\cC_{\rm PM}\le x$ has $P_{\rm PM}\le P_{\max}(x)$, where $P_{\max}$ is the inverse of $\cC^{(d)}_{\min}$ for $0\le x\le \cC_Q(d)$ and $P_{\max}(x)=P_Q(d)$ for $x\ge \cC_Q(d)$. For qubits, $\cC_C(2)=2[h_2(\tfrac34)-h_2(\tfrac12+\tfrac1{2\sqrt2})]\simeq0.421$ bits. The classical strategy has total irrealism one bit at $P=3/4$, more than $\cC_C(2)$ and without advantage: the bound~\eqref{eq:classval} is necessary, not sufficient.

The frontier also has an operational reading. Let $E$ be a system that purifies $\rho_{\cA\cB}$. Then $\Irr_B(\rho_{\cA\cB})=H(B|E)$~\cite{Coles2012}, the conditional entropy of the outcome of $B$ given $E$: since $B$ is rank one, measuring it on the pure state of $\cA\cB E$ leaves $\cA E$ in a pure state for each outcome $k$, so that $H(BE)=H(B)+\sum_kp_kS(\rho_\cA^{(k)})=S(\Phi_B(\rho_{\cA\cB}))$, while $H(E)=S(\rho_{\cA\cB})$. The irrealism is thus the uncertainty about the outcome of $B$ of an adversary who holds the purification of the preparation but not Alice's system. For a single system this is the interpretation of the relative entropy of coherence as intrinsic randomness against a quantum adversary~\cite{YuanZhao2016}, which extends the pure-state result of Ref.~\cite{Yuan2015}. In PM codes, a purification of the state~\eqref{eq:cq} consists of a copy of Alice's input $X$ and a system $E$ that purifies each $\rho_{ij}$, and the identity gives
\begin{equation}
\cC_{\rm PM}=H(b|EX)+H(b'|EX),
\label{eq:randomness}
\end{equation}
where $b$ and $b'$ are the outcomes of $B$ and $B'$; the adversary knows Alice's input and holds $E$. The frontier~\eqref{eq:frontier} is therefore the least value of this conditional entropy compatible with the success probability alone, a prepare-and-measure counterpart of source-independent randomness generation~\cite{Vallone2014,Cao2016,Marangon2017,Ma2019} and of entropy certification with semi-device-independent measurements~\cite{Li2011,WoodheadPironio2015,Brask2017,VanHimbeeck2017}. The statement holds under explicit assumptions: Bob's measurements are trusted and exactly mutually unbiased (Sec.~\ref{sec:decoders} relaxes this for qubits), his system is $d$-dimensional, and the adversary attacks each round independently; the von Neumann entropy then quantifies extractable randomness asymptotically for independent and identically distributed rounds, and classical mixtures of strategies are covered by the convexity of the frontier. With trusted measurements and a known dimension the states could also be found by tomography; the frontier is what the success probability alone certifies, and constraining the full statistics gives larger values. Whereas source-independent schemes bound the randomness through uncertainty relations or coherence witnesses~\cite{Cao2016,Ma2019}, the frontier gives its exact least value, in every dimension, as a function of the score of the code alone. Since Bob measures one decoding observable per round, chosen uniformly, the conditional entropy of his outcome per round is $\cC_{\rm PM}/2$; for a nonuniform input distribution the bound holds with $P$ the weighted success probability, by Jensen's inequality with weights.

The optimal code also admits an accounting of its irrealism. Since $\rho_{ij}(\theta)$ is the same quadratic polynomial of $M_{ij}$ for every $(i,j)$, and $\sum_{ij}M_{ij}^k\propto\id$ for $k=0,1,2$, Bob's average state is $\id/d$ at every $\theta$. For Bob's system alone both decoding observables are then elements of reality, and in Eq.~\eqref{eq:BA}, applied to the state~\eqref{eq:cq} with the input register $X$ in the role of $\cA$, only the correlation term survives. That term is the Holevo quantity $\chi_{\rm H}(X{:}\cB)$~\cite{Holevo1973} minus the classical mutual information between the input and the outcome, so that
\begin{equation}
\cC_{\rm PM}=2\chi_{\rm H}(X{:}\cB)-I(X{:}b)-I(X{:}b')
\label{eq:holevo}
\end{equation}
for every code whose average state is $\id/d$, in particular for the optimal codes: relative to Alice's input, the average local coherence equals twice the information that the code holds about the input minus the information that the two decoders extract.

It remains to see how the minimum at the classical value and the quantum endpoint scale with dimension. Both grow as $\log_2d$ (Appendix~\ref{app:asym}): $\cC_Q(d)$ exceeds $\log_2d$ by two bits up to corrections of order $\log_2d/\sqrt d$, whereas $\cC_C(d)$ approaches $\log_2d$ from below, with a deficit of order $(\ln d)^2/d$. The increment $\cC_Q-\cC_C$, which separates the best quantum code from the least unreal code at the classical value, equals $0.78$ bits for $d=2$, $1.15$ for $d=3$ and $1.37$ for $d=4$, and tends to two bits from below, though not monotonically [inset of Fig.~\ref{fig:dim}(c)].

The two bits have a simple origin. For the pure optimal state $S(\rho)=0$, so the irrealism of each decoding observable is the entropy of its outcome distribution, which is $P_Q$ on the intended outcome and uniform on the other $d-1$: $h_2(P_Q)+(1-P_Q)\log_2(d-1)\simeq1+\tfrac12\log_2d$. The two observables together give $\log_2d+2$, whereas the classical strategy, with $B$ real and $B'$ uniform, has total irrealism $\log_2d$. Since $\log_2d-\cC_C\to0$, for large $d$ the minimum at the classical value approaches the total irrealism of the classical strategy. The increment accumulates close to the quantum optimum, where the frontier is much sharper than at $P_C$ (Appendix~\ref{app:asym}). The exclusion bound behaves differently. Since $P_C$ and $P_Q$ both tend to $\tfrac12$, $2K_d\simeq\log_2d-2$ at both endpoints, so for large $d$ the difference from the exact minimum tends to two bits at the classical value and to four bits at the quantum optimum.

\subsubsection{Dependence on the decoders}\label{sec:decoders}

The frontier refers to mutually unbiased decoders. Minimizing Eq.~\eqref{eq:question} over the decoders as well gives a statement that does not depend on them.

\emph{Result (any decoders).} For PM $2_d\to1$ codes with rank-one projective decoders, the least total irrealism over all codes and all pairs of decoders with $P_{\rm PM}\ge P$ vanishes for $P\le P_C(d)$ and is strictly positive for $P_C(d)<P\le P_Q(d)$.

With $B'=B$, the code $\rho_{ij}=\Pi_i$ always answers the question about $i$, and answers the question about $j$ when $i=j$, so that $P_{\rm PM}=P_C(d)$ with no irrealism. Conversely, pairs of bases and codes form compact sets on which $\cC_{\rm PM}$ and $P_{\rm PM}$ are continuous, so for $P>P_C(d)$ the minimum is attained; if it vanished, $B$ would be an element of reality for every transmitted state, and the argument of Sec.~\ref{sec:PM} would give $P_{\rm PM}\le P_C(d)$. The zero below $P_C(d)$ is attained only by compatible decoders. A code with zero irrealism consists of states diagonal in both bases; for such a state $\Tr[\rho(\Pi_i+\Pi'_j)]\le1+|\langle b_i|b'_j\rangle|^2$, with equality, when $\langle b_i|b'_j\rangle\neq0$, only for $\rho=\Pi_i$, which must then be diagonal in the eigenbasis of $B'$ as well. Averaging over messages gives $P_{\rm PM}\le P_C(d)$, with equality only if every $\ket{b_i}$ is an eigenvector of $B'$. For a fixed pair of incompatible decoders, codes with zero irrealism therefore stay strictly below $P_C(d)$, and compactness makes the least irrealism strictly positive already at $P_C(d)$.

How fast the least irrealism rises beyond the classical value depends on the decoders. For qubits the closure property holds for every pair of decoding bases, because a dephased qubit state lies in the span of $\id$ and $\Pi_i$, and Eq.~\eqref{eq:anydec} gives the minimum exactly for decoders whose Bloch axes make an angle $\gamma$ (Appendix~\ref{app:decoders}). Minimized over $\gamma$ [Fig.~\ref{fig:dim}(a)], it vanishes for $P\le3/4$, where it is attained by $B'=B$, and lies strictly below the MUB frontier for $3/4<P<P_*$, with $P_*\simeq0.833$, because MUBs are not a local minimum over decoders there; at $P=0.76$ it is $0.134$ bits, against $0.463$ for MUBs. For $P_*<P<P_Q$, MUBs are a strict local minimum over decoders, and a numerical minimization over $\gamma$ indicates that they are the global minimum; at $P_Q$ only MUBs reach the success probability~\cite{Farkas2019}. The origin of the difference is visible message by message. The decoding operator of message $(i,j)$ is $\Pi_i+\Pi'_j=\id+\bm m_{ij}\cdot\bm\sigma$, with $|\bm m_{ij}|=\cos(\gamma/2)$ for $i=j$ and $|\bm m_{ij}|=\sin(\gamma/2)$ for $i\neq j$, and the least irrealism at a given success probability of the message above $\tfrac12$ strictly decreases with $|\bm m_{ij}|$. Decreasing $\gamma$ below $\pi/2$ therefore lowers the cost of the messages with $i=j$, raises that of the messages with $i\neq j$, and lowers the largest attainable success probability, $\tfrac12+\tfrac14[\cos(\gamma/2)+\sin(\gamma/2)]$. Below $P_*$ the saving outweighs the extra cost; above $P_*$ it does not, and mutually unbiased decoders, which are the most incompatible, are at least locally also the least costly.

Nearly aligned decoders allow an advantage with little irrealism, but only a small advantage: as $\gamma\to0$ the largest success probability tends to $3/4$. The irrealism per unit of advantage does not vanish. 

The same analysis makes the certification of Sec.~\ref{sec:gibbs} robust for qubits. If the decoders are only known to make an angle within $\delta$ of $\pi/2$, an observed success probability $P$ certifies $\min_{|\gamma-\pi/2|\le\delta}\cC_{\min}(P\,|\,\gamma)$, which Eq.~\eqref{eq:anydec} gives. For $P_*<P<P_Q$ the strict local minimality of mutually unbiased decoders implies that this equals the MUB frontier for $\delta$ small enough; below $P_*$ it is smaller. For $d>2$ the dependence on the decoders is open.

\subsection{Entanglement-assisted codes}\label{sec:EA}

\begin{figure*}[t]
\centering
\includegraphics[width=\textwidth]{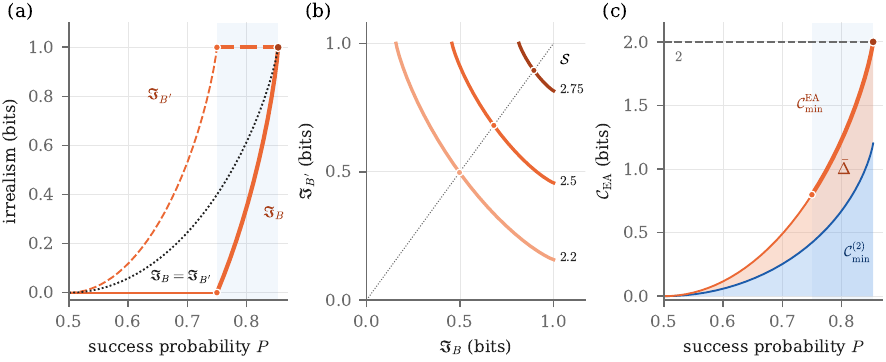}
\caption{Irrealism of entanglement-assisted codes, with Bob decoding in $B=\sigma_z$ and $B'=\sigma_x$, and $\cS=8P-4$. Thin and thick lines and dots as in Fig.~\ref{fig:dim}; the light band in (a) and (c) is the quantum-advantage region $P>3/4$. (a) Irrealism of each decoding observable in two kinds of optimal codes. Solid: the minimum of $\Irr_B$, Eq.~\eqref{eq:threshold}, which leaves zero linearly at $P=3/4$. Dashed: the least $\Irr_{B'}$ compatible with it by Eq.~\eqref{eq:tradeoff}, equal to $\varphi(\cS/2)$ for $\cS\le2$ and to $1$ for $\cS>2$, attained by $\rho_{\cS/2,0}$ and $\rho_{1,t_x}$, respectively. Dotted: $\Irr_B=\Irr_{B'}=\varphi(\cS/2\sqrt2)$ in the codes that minimize the total, Eq.~\eqref{eq:EAcost}. Up to $P=3/4$, $B$ can be made real by moving all the irrealism to $B'$; beyond it, $B'$ is already maximally unreal and $B$ cannot be real. The three curves meet at one bit at the Tsirelson point. (b) Boundary of the region allowed by the irrealism--CHSH trade-off, Eq.~\eqref{eq:tradeoff}, for $\cS=2.2$, $2.5$, $2.75$; the allowed region lies above and to the right of each curve, every point on the curves is attained by the states~\eqref{eq:extremal}, and the dots, on the diagonal, mark the minimum total irrealism at each $\cS$. (c) Exact frontier $\cC^{\rm EA}_{\min}$, Eq.~\eqref{eq:EAcost} (orange line), with the upper bound of two bits (dashed), which follows from $\Irr_B,\Irr_{B'}\le1$, decomposed as in Eq.~\eqref{eq:steerbound}: the blue region is the PM frontier $\cC^{(2)}_{\min}$ of the steered code, Eq.~\eqref{eq:qubit}, and the orange region, of height $\bar\Delta=2[1-h_2(P)]$, Eq.~\eqref{eq:gap}, is the reality variation that Alice's measurements induce on the decoding observables, that is, the irrealism that her measurement turns into information relative to her outcome (Sec.~\ref{sec:EA}). The window closes at the Tsirelson point, with value $2$ bits. All curves follow from closed-form expressions.}
\label{fig:main}
\end{figure*}

In the EA code the information Bob needs is not in his qubit, whose marginal can be maximally mixed, but in the correlations with Alice, and Eq.~\eqref{eq:exclusion2} charges it to the irrealism of Bob's other observable. Through $\Irr_B=H(B|E)$, the bounds that follow are entropy bounds of the one-sided device-independent type, with Bob's measurements trusted and Alice's uncharacterized~\cite{Branciard2012}. As there, we assume an exact qubit on Bob's side measured in $\sigma_z$ and $\sigma_x$, an adversary holding a purification $E$ of $\rho_{\cA\cB}$, collective attacks on independent and identically distributed rounds, and Alice's setting depending only on $z$. The bound on each irrealism is the trusted-qubit form of known device-independent bounds~\cite{Acin2007,Pironio2009,Masini2022}; read through the same identity, the trade-off and the total minimum below are uncertainty relations for Bob's two observables certified by the CHSH value, in the spirit of Refs.~\cite{Kaniewski2014,TomamichelHanggi2013}. What we add is the minimum of the total irrealism, its tightness for any system on Alice's side, and its relation to the PM frontier.

For a qubit on Alice's side, let $\bm u$ and $\bm v$, with $u_k=\Tr[\rho\,\sigma_k\otimes B]$ and $v_k=\Tr[\rho\,\sigma_k\otimes B']$, be Alice's correlation vectors with $B$ and $B'$, so that $\bm u=T\hat z$ and $\bm v=T\hat x$ with $T_{kl}=\Tr[\rho\,\sigma_k\otimes\sigma_l]$. For traceless observables $A=\bm a\cdot\bm\sigma$ and $A'=\bm a'\cdot\bm\sigma$ with $|\bm a|,|\bm a'|\le1$, $\cS=\bm a\cdot(\bm u+\bm v)+\bm a'\cdot(\bm u-\bm v)\le2(|\bm u|^2+|\bm v|^2)^{1/2}$, as in the Horodecki criterion~\cite{Horodecki1995}. The exclusion bounds the same lengths. With
\begin{equation}
\varphi(t)=1-h_2\!\Big(\frac{1+t}{2}\Big)=\frac1{\ln2}\sum_{k\ge1}\frac{t^{2k}}{2k(2k-1)},
\label{eq:phi}
\end{equation}
the information carried by a binary variable with bias $t\in[0,1]$, whose series shows that $\varphi$ and $\chi(w)\equiv\varphi(\sqrt w)$ are convex and increasing, Alice's measurement of $\hat{\bm v}\cdot\bm\sigma$ in Eq.~\eqref{eq:exclusion2} gives $\Irr_B\ge\varphi(|\bm v|)$, and likewise $\Irr_{B'}\ge\varphi(|\bm u|)$, the qubit form of the bound on an eavesdropper's uncertainty in terms of the error in the complementary basis~\cite{Tomamichel2012,Masini2022}. Since $\Irr_B,\Irr_{B'}\le1$ for a qubit (Appendix~\ref{app:EA}), combining the two bounds gives the trade-off
\begin{equation}
\big[\varphi^{-1}\big(\Irr_B(\rho_{\cA\cB})\big)\big]^2+\big[\varphi^{-1}\big(\Irr_{B'}(\rho_{\cA\cB})\big)\big]^2\ \ge\ \frac{\cS^2}{4}.
\label{eq:tradeoff}
\end{equation}
It holds for any quantum system on Alice's side and any two-outcome measurements of hers: a Naimark dilation and Jordan's lemma reduce her observables to blocks of dimension at most two, a dephasing onto these blocks leaves $\cS$ unchanged and cannot increase either irrealism, and the left-hand side of Eq.~\eqref{eq:tradeoff}, being concave in the two irrealisms, combines the blocks (Appendix~\ref{app:EA}). Every point of the boundary is attained by the Bell-diagonal states
\begin{equation}
\rho_{t_z,t_x}=\tfrac14\big(\id+t_x\,\sigma_x\otimes\sigma_x-t_xt_z\,\sigma_y\otimes\sigma_y+t_z\,\sigma_z\otimes\sigma_z\big),
\label{eq:extremal}
\end{equation}
with $t_z,t_x\in[0,1]$, for which $\Irr_B=\varphi(t_x)$, $\Irr_{B'}=\varphi(t_z)$, $\cS=2(t_z^2+t_x^2)^{1/2}$, and $\rho_\cA=\rho_\cB=\id/2$, so that the irrealism is entirely the correlation term of Eq.~\eqref{eq:BA}; for a given $\cS>2$ the boundary points have $t_z,t_x\ge(\cS^2/4-1)^{1/2}$. The irrealism of $B=\sigma_z$ is set by $t_x$ because an unrevealed measurement of $\sigma_z$ on Bob's side removes the $\sigma_x\otimes\sigma_x$ and $\sigma_y\otimes\sigma_y$ correlations and keeps $\sigma_z\otimes\sigma_z$. Once Bell violation is demanded, making one decoding observable more real forces the other to be less real [Fig.~\ref{fig:main}(b)]. Because $\varphi^{-1}\le1$, Eq.~\eqref{eq:tradeoff} also bounds each observable separately,
\begin{equation}
\Irr_B(\rho_{\cA\cB}),\ \Irr_{B'}(\rho_{\cA\cB})\ \ge\ \varphi\Big(\sqrt{\max\{\cS^2/4-1,0\}}\Big),
\label{eq:threshold}
\end{equation}
with equality for $\rho_{1,t_x}$ and $\rho_{t_z,1}$ at $t_{x},t_z=(\cS^2/4-1)^{1/2}$. Through $\Irr_B=H(B|E)$, Eq.~\eqref{eq:threshold} is the device-independent bound on an eavesdropper's uncertainty as a function of the CHSH value~\cite{Acin2007,Pironio2009}, here with fixed decoders. Near $\cS=2$, $\cS^2/4-1\simeq\cS-2=8(P_{\rm EA}-\tfrac34)$ and $\varphi(t)\simeq t^2/(2\ln2)$, so the threshold leaves zero linearly, with slope $4/\ln2$, and it reaches one bit at the Tsirelson point. If either decoding observable is real, then $P_{\rm EA}\le3/4$; this holds for any rank-one projective decoders, because a real $B$ makes $\rho_{\cA\cB}=\sum_kp_k\rho_\cA^{(k)}\otimes\proj{b_k}$ separable. As in PM codes (Sec.~\ref{sec:PM}), a decoding observable can therefore be real exactly up to the classical value [Fig.~\ref{fig:main}(a)].

\begin{table*}[t]
\caption{Irrealism frontier at success probability $P$. In all EA results Alice's setting depends only on $z$, her measurements have two outcomes, and Bob decodes with $B=\sigma_z$, $B'=\sigma_x$. That a real decoding observable forbids an advantage holds for arbitrary rank-one projective decoders in both architectures. For arbitrary qubit decoders at angle $\gamma$, the minimum over $\gamma$ is evaluated numerically.}
\label{tab:results}
\begin{ruledtabular}
\begin{tabular}{lllll}
Quantity & Code & Status & Values & Eq.\\
\hline
One observable & PM, any $d$, MUBs & one-parameter reduction & $0$ for $P\le P_C$; $\cC_Q/2$ at $P_Q$ & \eqref{eq:seqoned}\\
 & PM, any $d$, MUBs & bound & $K_d(2P-1)$ & \eqref{eq:Kd}\\
 & EA, any $\cA$ & exact; DI bound~\cite{Acin2007,Pironio2009} & $0$ for $P\le\tfrac34$; $1$ at $\cS=2\sqrt2$ & \eqref{eq:threshold}\\
 & EA, any $\cA$ & trade-off, tight & attained, Eq.~\eqref{eq:extremal} & \eqref{eq:tradeoff}\\
Both observables & PM, any $d$, MUBs & exact, unique optimum & $\cC_C(2)=0.42$, $\cC_Q(2)=1.20$; $\cC_Q-\cC_C\to2$ & \eqref{eq:frontier}\\
 & PM, any $d$, MUBs & bound (exclusion) & $2K_d(P)$ & \eqref{eq:exbound}\\
 & PM, any $d$, MUBs & upper bound, all MUB codes & $2[h_2(P)+(1-P)\log_2(d-1)]$; $\cC_Q$ at $P_Q$ & \eqref{eq:window}\\
 & EA, any $\cA$ & exact & $0.80$ at $P=\tfrac34$; $2$ at $\cS=2\sqrt2$ & \eqref{eq:EAcost}\\
 & EA, any $\cA$ & bound, steered PM code & $\cC^{(2)}_{\min}(P)+\bar\Delta$ & \eqref{eq:steerbound}\\
 & PM, $d=2$, any decoders & exact at fixed $\gamma$ & $\min_\gamma=0$ up to $\tfrac34$; MUBs beaten below $P_*$ & \eqref{eq:anydec}\\
\end{tabular}
\end{ruledtabular}
\end{table*}

\subsubsection{Exact frontier}\label{sec:EAcost}

For the total irrealism the natural constraint is the sum relation~\eqref{eq:memory}. For a qubit on Alice's side, let $t_{(1)}\ge t_{(2)}\ge t_{(3)}$ be the singular values of $T$. Then
\begin{equation}
\cC_{\rm EA}\ \ge\ 1-S(\cB|\cA)_\rho\ \ge\ \varphi(t_{(1)})+\varphi(t_{(2)}).
\label{eq:EAsv}
\end{equation}
The first inequality is Eq.~\eqref{eq:memory} with $d_\cB=2$. For the second, local rotations bring $T$ to diagonal form and a local Pauli twirl removes the local Bloch vectors without changing $T$, which by the concavity of the conditional entropy cannot decrease $S(\cB|\cA)$; the resulting Bell-diagonal state is a distribution of two bits whose biases are two diagonal entries of $T$, and subadditivity gives the bound (Appendix~\ref{app:EA}). Every inequality in Eq.~\eqref{eq:EAsv} is an equality for the states~\eqref{eq:extremal}, whose two largest singular values are $t_z$ and $t_x$. The success probability enters through $\cS^2/4\le|\bm u|^2+|\bm v|^2\le t_{(1)}^2+t_{(2)}^2$, since $\bm u$ and $\bm v$ are the images under $T$ of two orthonormal vectors, and the convexity and monotonicity of $\chi$ give
\begin{equation}
\cC_{\rm EA}\ \ge\ 2\varphi\Big(\frac{\cS}{2\sqrt2}\Big)=2\Big[1-h_2\Big(\frac12+\frac{2P_{\rm EA}-1}{\sqrt2}\Big)\Big]
\label{eq:EAcost}
\end{equation}
for $\cS\ge0$; it also holds for $\cS<0$, since relabeling Alice's outcomes reverses the sign of $\cS$ and leaves the irrealisms unchanged. Equality holds for the Gibbs resource $\rho_{t,t}=e^{\beta W}/\Tr e^{\beta W}$, with $W=\sqrt2(\sigma_z\otimes\sigma_z+\sigma_x\otimes\sigma_x)$ and $t=\tanh(\sqrt2\,\beta)=\cS/(2\sqrt2)$, when Alice measures $A=(\sigma_z+\sigma_x)/\sqrt2$ and $A'=(\sigma_z-\sigma_x)/\sqrt2$. Appendix~\ref{app:EA} removes the assumptions of a qubit on Alice's side and traceless observables: after the block reduction used for Eq.~\eqref{eq:tradeoff}, the two-dimensional blocks obey the qubit bound, and in the one-dimensional blocks, where Alice's outcomes are fixed, the irrealism of Bob's qubit alone exceeds it. Equation~\eqref{eq:EAcost} therefore holds for any quantum system on Alice's side and any two-outcome measurements, and its right-hand side is the EA frontier $\cC^{\rm EA}_{\min}(P)$. Every EA code with a quantum advantage has $\cC_{\rm EA}>2\varphi(1/\sqrt2)\simeq0.798$ bits, and since each irrealism is at most one bit, $2\varphi(\cS/2\sqrt2)\le\cC_{\rm EA}\le2$; this window closes at the Tsirelson bound, where both decoding observables are maximally unreal, in accordance with the rigidity of the CHSH inequality~\cite{Supic2020}.

Through $\Irr_B=H(B|E)$, Eq.~\eqref{eq:EAcost} is a closed-form lower bound on $H(B|E)+H(B'|E)$ in terms of the CHSH value with Bob's decoders trusted, the trusted-decoder counterpart of the two-basis analysis of Ref.~\cite{Masini2022}. There, the per-observable bound used above appears as a qubit uncertainty relation, and the two-basis bound rests on the convexity of $\chi$ and on Jordan's lemma; with the trusted constraint $|\bm u|^2+|\bm v|^2\ge\cS^2/4$ these tools give Eq.~\eqref{eq:EAcost} for two qubits directly. Its right-hand side is also the BB84 term $2[1-h_2(Q)]$ at the error rate $Q=\tfrac12-\cS/(4\sqrt2)$. Without trusted decoders the problem underlies device-independent key distribution with a random key basis, bounded numerically~\cite{Schwonnek2021} and analytically~\cite{Masini2022}, the latter building on the analytical bound for a single key basis and asymmetric CHSH expressions of Ref.~\cite{Woodhead2021}; there the same expression appears only as the value of an explicit attack which, convexified with a deterministic strategy, is conjectured optimal from numerics~\cite{Masini2022}. The device-independent minimum is convex in $\cS$, since mixing strategies with a classical flag averages both the CHSH value and the entropies; it vanishes at $\cS=2$ and is at most two bits at $\cS=2\sqrt2$, so it lies below the chord $2(\cS-2)/(2\sqrt2-2)$, and hence strictly below Eq.~\eqref{eq:EAcost}, for $2\le\cS<2.695$. Mixing the attack with a deterministic strategy at $\cS=2$ gives the convex envelope of the point $(2,0)$ and the curve~\eqref{eq:EAcost}, which coincides with that curve for $\cS\ge\cS_t\simeq2.776$, i.e., $P_{\rm EA}\gtrsim0.847$. The device-independent minimum is therefore strictly below Eq.~\eqref{eq:EAcost} for $2\le\cS<\cS_t$, and if the convexified attack is optimal, trusting Bob's decoders gains nothing for $\cS\ge\cS_t$. The trusted setting provides the proof that Eq.~\eqref{eq:EAcost} is a tight lower bound for any system on Alice's side, the trade-off~\eqref{eq:tradeoff} in the same generality, and the comparison with the PM frontier below.

The comparison is an accounting, because the two frontiers are evaluated relative to different records: $\cC_{\rm EA}$ on the resource before Alice measures, $\cC_{\rm PM}$ relative to Alice's input. When Alice measures $A_z$ and obtains $a$, she steers Bob's qubit into $\rho^{a|z}_\cB$ with probability $p(a|z)$, and the ensemble $\{p(a|z)/2,\rho_\cB^{a|z}\}$ is a PM code with message $(i,j)=(a,a\oplus z)$ and success probability $P_{\rm EA}$. The steered code is a remote preparation rather than a PM protocol: no quantum system is transmitted, Alice does not choose the state that Bob receives, and the message $(a,a\oplus z)$ is fixed by her outcome; it is an ensemble to which the PM frontier applies as a bound. With $\mathcal M_z(\rho_{\cA\cB})=\sum_a\proj a\otimes p(a|z)\rho_\cB^{a|z}$ her measurement with the outcome recorded but not revealed, for any shared state and any two-outcome measurements of Alice,
\begin{equation}
\Irr_B(\rho_{\cA\cB})=\sum_ap(a|z)\,\Irr_B(\rho^{a|z}_\cB)+\Delta^z_B,
\label{eq:steer}
\end{equation}
where $\Delta^z_B=\Irr_B(\rho_{\cA\cB})-\Irr_B(\mathcal M_z(\rho_{\cA\cB}))\ge0$ by data processing under a map on Alice's side, which commutes with $\Phi_B$, and likewise for $B'$. Averaging over $z$ and applying the PM frontier to the steered code, whose message weights $p(a|z)/2$ need not be uniform, gives, for $P_{\rm EA}\ge\tfrac12$ (Appendix~\ref{app:EA}),
\begin{equation}
\cC_{\rm EA}\ \ge\ \cC^{(2)}_{\min}(P_{\rm EA})+\bar\Delta,\qquad \bar\Delta=\frac12\sum_z\big(\Delta^z_B+\Delta^z_{B'}\big)\ge0 .
\label{eq:steerbound}
\end{equation}
For a rank-one projective measurement of Alice, the purification identity applied before and after the measurement gives (Appendix~\ref{app:EA})
\begin{equation}
\Delta^z_B=H(B|E)-H(B|E,a)=I(B{:}a|E),
\label{eq:CMI}
\end{equation}
the information about the outcome of $B$ that Alice's outcome adds to that held by a purifying system; for a general measurement of Alice, $\Delta^z_B\ge I(B{:}a|E)$ (Appendix~\ref{app:EA}). For a rank-one projective measurement, $\mathcal M_z(\rho_{\cA\cB})$ coincides with $\Phi_{A_z}(\rho_{\cA\cB})$ up to a relabeling of Alice's basis, so that $\Delta^z_B=\eta_{B|A_z}(\rho_{\cA\cB})$, the realism-based nonlocality of the context $(A_z,B)$, Eq.~\eqref{eq:rbn}. By the symmetry of Eq.~\eqref{eq:rbn}, $\Delta^z_B=\eta_{A_z|B}(\rho_{\cA\cB})$ is also the reality that Bob's decoding measurement confers on Alice's observable, so the accounting does not presuppose a temporal order between the two measurements. Since $\Tr_\cA\mathcal M_z(\rho_{\cA\cB})=\rho_\cB$, the reality variation changes only the correlation term of Eq.~\eqref{eq:BA}, as no-signaling requires. For the optimal resource the steered states are $\rho_\cB^{a|z}=\tfrac12[\id+(-1)^at\,\hat n_z\cdot\bm\sigma]$, with $\hat n_{0,1}=(\hat z\pm\hat x)/\sqrt2$, $p(a|z)=\tfrac12$, and $t=\sqrt2(2P-1)$: they are the optimal PM states at the same $P$, Eq.~\eqref{eq:steerbound} is an equality, and $\Delta^z_B=\Delta^z_{B'}=1-h_2(P)$, the information about each decoding observable in the steered state. Hence
\begin{equation}
\cC^{\rm EA}_{\min}(P)-\cC^{(2)}_{\min}(P)=2\big[1-h_2(P)\big],
\label{eq:gap}
\end{equation}
the orange region of Fig.~\ref{fig:main}(c), which for qubits equals the exclusion bound $2K_2(P)$: relative to her outcome, Alice's measurement turns irrealism of each decoding observable into the same amount of information about it. Given Eqs.~\eqref{eq:qubit} and~\eqref{eq:EAcost}, Eq.~\eqref{eq:gap} is a direct subtraction; its content lies in the tightness of Eq.~\eqref{eq:steerbound} for the optimal code above and in the identification of $\bar\Delta$. The excess of the EA frontier over the PM one is thus the realism-based nonlocality of the four contexts $(A_z,B)$ and $(A_z,B')$ used by the optimal code, summed over the two decoders and averaged over Alice's settings. Each of the four terms equals $1-h_2(P)$ and is at most $\eta_{\max}(\rho_{\cA\cB})$, so the optimal resource has $\eta_{\max}\ge1-h_2(P)$.

The same decomposition bounds the realism-based nonlocality from below for every code in which Alice's measurements are rank-one projective. Averaged over $z$, Eq.~\eqref{eq:steer} is an exact accounting, $\cC_{\rm EA}=\cC^{\rm st}_{\rm PM}+\bar\Delta$, with $\cC^{\rm st}_{\rm PM}$ the total irrealism of the steered code, and for such measurements each of the four terms of $\bar\Delta$ is again at most $\eta_{\max}(\rho_{\cA\cB})$. Bounding $\cC_{\rm EA}$ from below by the EA frontier, Eq.~\eqref{eq:EAcost}, and $\cC^{\rm st}_{\rm PM}$ from above by the upper bound~\eqref{eq:window} applied to the steered code gives (Appendix~\ref{app:EA})
\begin{equation}
\eta_{\max}(\rho_{\cA\cB})\ \ge\ \varphi\big(\sqrt2(2P_{\rm EA}-1)\big)-h_2(P_{\rm EA}).
\label{eq:rbncert}
\end{equation}
The right-hand side is half the difference between the least total irrealism that the success probability requires of the shared state and the largest that the steered code can carry. It vanishes at $P_{\rm EA}\simeq0.8145$, is positive above this value, and reaches $1-h_2(P_Q)\simeq0.399$ bits at the Tsirelson point. Under the assumptions of this section, with trusted decoders and Alice's measurements rank-one projective, an observed success probability above this threshold therefore certifies a minimum realism-based nonlocality of the shared state. A Bell violation already certifies entanglement; the content of Eq.~\eqref{eq:rbncert} is the quantitative bound. The bound is not tight, because Eq.~\eqref{eq:window} bounds the steered code by its worst case and $\eta_{\max}$ is bounded from below only by an average over four contexts: at the Tsirelson point the bound gives the value of each context of the optimal code, whereas the shared state is maximally entangled~\cite{Supic2020}, with $\eta_{\max}=1$~\cite{Gomes2018}.

Table~\ref{tab:results} collects the main results of Secs.~\ref{sec:PM} and~\ref{sec:EA} and their assumptions.

\section{Conclusion}\label{sec:conclusion}

We have determined how much irrealism the decoding observables of a quantum random access code must have at each success probability. For any rank-one projective decoders, a code in which either decoding observable is an element of reality cannot beat the classical value, and in PM codes the least total irrealism, minimized also over the decoders, vanishes exactly up to the classical value and is positive beyond it. For decoders in mutually unbiased bases, the least total irrealism of PM $2_d\to1$ codes is known exactly in every dimension, attained uniquely by Gibbs codes, and lies strictly above the exclusion bound; an upper bound obeyed by every code meets it at the quantum optimum, where the success probability fixes the irrealism, $\cC_Q(d)\simeq\log_2d+2$ bits. For entanglement-assisted qubit codes with trusted decoders, the closed-form minimum holds for any system on the sender's side with two-outcome measurements and exceeds the PM frontier by $2[1-h_2(P)]$, the information about the decoding observables that the sender's outcome provides, which is the realism-based nonlocality of the contexts used by the optimal code, summed over the two decoders; with rank-one projective sender measurements, a success probability above $0.815$ certifies a minimum realism-based nonlocality of the shared state.

The method behind the PM frontier is the excess identity~\eqref{eq:excess}: a closure property of the unrevealed measurements makes a stationary point the unique global minimum. It uses mutual unbiasedness through the equality of the overlaps $|\langle b_k|b'_j\rangle|$. For generic bases $\Phi_B(e^{\theta M})$ need not lie in the span of $\id$ and $\Pi_i$, and the Gibbs construction need not remain optimal, although any full-rank optimum must satisfy Eq.~\eqref{eq:closure}. For qubits, closure holds for every pair of decoders and the minimum is known exactly; mutually unbiased decoders are not the cheapest below $P_*\simeq0.833$, and above it they are a strict local minimum, which a numerical minimization indicates is global.

Several extensions remain open: the role of the decoders for $d>2$; untrusted decoders, which would combine the frontier with robust self-testing of mutually unbiased bases~\cite{Farkas2019}; noisy decoders, $n\to1$ codes, constraints on the full statistics, and certification beyond independent and identically distributed rounds; EA codes in which Alice's setting depends on both bits~\cite{Kanjilal2023} or Bob holds a qudit, where the two architectures differ~\cite{Tavakoli2016}; and the analogous question for the oracle-problem communication tasks of Ref.~\cite{Teeni2026}.

In the sense of the criterion~\eqref{eq:criterion}, a quantum advantage in these codes requires decoding observables without definite values, and the information--reality exclusion gives this requirement the form of a complementarity: the information that a code delivers about one decoding observable is paid for by the irrealism of the other, and at the quantum optimum the two are locked, since the performance alone fixes the irrealism. Entanglement-assisted codes add a relational aspect, in the spirit of Bohr's reply to EPR. The same decoding observables are unreal relative to the shared resource, maximally so at the Tsirelson point, while relative to Alice's outcome their irrealism is that of a PM code; the difference is the reality variation that her distant measurement induces without changing Bob's local state, the realism-based nonlocality of the contexts that the code uses~\cite{Bilobran2015}. Whether a decoding observable has a definite value thus depends on the whole experimental arrangement, including measurements on a distant system whose outcomes are not revealed, and the frontiers quantify the least indefiniteness that each arrangement requires for a given performance.

\begin{acknowledgments}
The author thanks Ismael L. Paiva, Breno Marques, and Renato M. Angelo for useful comments. P.R.D. acknowledges support from CNPq/MCTI/FNDCT 22/2024 (No. 446775/2024-0).
\end{acknowledgments}

\bibliography{irrealism_qrac}

\appendix

\section{One decoding observable in prepare-and-measure codes}\label{app:seqone}

\emph{Bound~\eqref{eq:Kd}.} The exclusion~\eqref{eq:exclusion}, applied to a single transmitted state, gives $\Irr_B(\rho_{ij})\ge\Info_{B'}(\rho_{ij})=\log_2d-H_{B'}(\rho_{ij})$: the irrealism of $B$ is at least the information that the state carries about $B'$. If the intended outcome $j$ of $B'$ has probability $q_{ij}=\Tr\rho_{ij}\Pi'_j$, the entropy $H_{B'}$ is largest when the remaining probability $1-q_{ij}$ is spread uniformly over the other $d-1$ outcomes, so that $\Irr_B(\rho_{ij})\ge K_d(q_{ij})$, with $K_d$ the function of Eq.~\eqref{eq:Kdef}. To express this through the success of the message, note that $q_{ij}=2s_{ij}-p_{ij}\ge2s_{ij}-1$. The function $K_d$ is convex, with $K_d'(q)=\log_2\frac{q(d-1)}{1-q}$, and vanishes together with its derivative at $q=1/d$, so it increases for $q\ge1/d$. Hence $\Irr_B(\rho_{ij})\ge k_d(s_{ij})$, where $k_d(s)=K_d(2s-1)$ for $s\ge P_C(d)$, i.e., $2s-1\ge1/d$, and $k_d(s)=0$ otherwise. Since the two pieces join with zero slope at $P_C(d)$, $k_d$ is convex and nondecreasing, and Jensen's inequality over messages, $d^{-2}\sum_{ij}k_d(s_{ij})\ge k_d(P_{\rm PM})$, gives Eq.~\eqref{eq:Kd}. Exchanging the roles of $B$ and $B'$, with $p_{ij}\ge2s_{ij}-1$, gives the same bound for $B'$.

\emph{Every dimension, Eqs.~\eqref{eq:seqoned} and~\eqref{eq:seqonep}.} The proof proceeds in five steps.

(i) \emph{Same problem for every message.} Fix $(i,j)$. Since $|\langle b_k|b'_j\rangle|=d^{-1/2}$ for every $k$, a unitary $V$ diagonal in the eigenbasis of $B$ maps $\ket{b'_j}$ to $\ket{+}=d^{-1/2}\sum_k\ket{b_k}$. Being diagonal in that basis, $V$ commutes with $\Phi_B$ and fixes $\Pi_i$, so it preserves $\Irr_B$ and $\Tr\rho\Pi_i$, while it carries $\Tr\rho\Pi'_j$ into $\bra{+}\rho\ket{+}$. A permutation of the eigenvectors of $B$, which fixes $\ket{+}$, changes $i$ in the same way. The single-message minimum $\tilde g(s)=\inf\{\Irr_B(\rho):\tfrac12\Tr[\rho(\Pi_i+\Pi'_j)]=s\}$ is therefore the same for all messages. From now on we write $\Pi'_j=\proj{+}$.

(ii) \emph{Symmetric states suffice.} The permutation unitaries $W$ of the eigenvectors $\{\ket{b_k}\}_{k\neq i}$ fix $\Pi_i$ and $\ket{+}$ and commute with $\Phi_B$, so they leave both the constraint and $\Irr_B$ unchanged. Because $\Irr_B(\rho)=D(\rho\Vert\Phi_B(\rho))$ is convex in $\rho$, replacing a state by its average over these permutations keeps the constraint and cannot increase $\Irr_B$; the infimum can thus be taken over permutation-invariant states. For these, the probabilities $\bra{b_k}\rho\ket{b_k}$ of the unintended outcomes $k\neq i$ are all equal, so the outcome distribution of $B$ is fixed by $p=\Tr\rho\Pi_i$ alone, $S(\Phi_B(\rho))=f_d(p)$, and $\Irr_B(\rho)=f_d(p)-S(\rho)$.

(iii) \emph{Maximum entropy at fixed decoding probabilities.} At fixed $p$ the first term is fixed, and what remains is to make $S(\rho)$ as large as possible at fixed $p$ and $q=\Tr\rho\Pi'_j$. The Gibbs states $\rho_{\lambda\mu}$ do this. Since $\log\rho_{\lambda\mu}$ is a combination of $\id$, $\Pi_i$ and $\Pi'_j$, every state $\sigma$ with the same $p$ and $q$ has $\Tr(\sigma\log\rho_{\lambda\mu})=\Tr(\rho_{\lambda\mu}\log\rho_{\lambda\mu})$, and therefore $S(\rho_{\lambda\mu})-S(\sigma)=D(\sigma\Vert\rho_{\lambda\mu})\ge0$. This is the quantum maximum-entropy principle~\cite{Wichmann1963,WeisKnauf2012}: every pair $(p,q)$ in the interior of the attainable set is reached by one state $\rho_{\lambda\mu}$, and pairs on its boundary by limits of them. The states $\rho_{\lambda\mu}$ are themselves permutation invariant, so the maximum is reached within the class of step (ii). With $S_{\max}(p,q)$ the largest entropy at fixed $(p,q)$ and $q=2s-p$, every invariant state obeys $\Irr_B(\rho)\ge f_d(p)-S_{\max}(p,2s-p)$, with equality for the Gibbs state. Minimizing over $p$ gives Eq.~\eqref{eq:seqonep}, and restricting to Gibbs states gives Eq.~\eqref{eq:seqoned}, a minimization over the single combination of $\lambda$ and $\mu$ left free by the constraint. Hence $\tilde g(s)$ is the right-hand side of Eq.~\eqref{eq:seqoned} at $P=s$.

(iv) \emph{Boundary and endpoint.} Equation~\eqref{eq:seqoned} runs only over Gibbs states with $(p,q)$ in the interior of the attainable set, and two points need care. First, the minimum over $p$ could lie on the boundary of that set, where the maximum-entropy state is only a limit of Gibbs states and can even jump as $(p,q)$ varies~\cite{WeisKnauf2012}. This does not matter, because Eq.~\eqref{eq:seqonep} uses only the value $S_{\max}(p,q)$, not the state. This value is upper semicontinuous in $(p,q)$, being the maximum of the continuous function $S$ over the compact set of states with the given expectation values; hence the minimum over $p$ in Eq.~\eqref{eq:seqonep} is attained. It is also concave, because $S$ is concave and the constraints are linear, and a concave, upper semicontinuous function on a segment is continuous on the whole segment, endpoints included. Along the segment $(p,2P-p)$ the interior values therefore approach the boundary ones, and the infimum over interior Gibbs states in Eq.~\eqref{eq:seqoned} equals the minimum in Eq.~\eqref{eq:seqonep}. Second, at $P=P_Q$ the only feasible state is the top eigenvector of $M_{ij}$, which is not of Gibbs form, and Eq.~\eqref{eq:seqoned} is read as the limit $P\to P_Q$.

(v) \emph{Properties of $\tilde g$ and the full code.} As the value function of a convex functional, $\Irr_B$, under a linear constraint, $\tilde g$ is convex. It vanishes exactly up to the classical value. A state with $\Irr_B=0$ is diagonal in the eigenbasis of $B$, so its outcome distribution for $B'$ is uniform and $q=1/d$, which gives $s=(p+1/d)/2\le P_C(d)$; since the minimum is attained, $\tilde g(s)>0$ for $s>P_C(d)$. Conversely, for $1/d\le s\le P_C(d)$ the diagonal states with $p=2s-1/d$ have zero irrealism, so $\tilde g(s)=0$ there. Being convex, zero on $[1/d,P_C(d)]$ and positive beyond, $\tilde g$ is nondecreasing on $[1/d,P_Q]$, so the constraints $P_{\rm PM}=P$ and $P_{\rm PM}\ge P$ give the same minimum. At $s=P_Q$ the only feasible state is the top eigenvector of $M_{ij}$, whose outcome distribution for $B$ is $P_Q$ on $i$ and uniform elsewhere, so $\tilde g(P_Q)=f_d(P_Q)=\cC_Q(d)/2$. For the whole code, Jensen's inequality over messages gives $d^{-2}\sum_{ij}\Irr_B(\rho_{ij})\ge d^{-2}\sum_{ij}\tilde g(s_{ij})\ge\tilde g(P_{\rm PM})$, with equality for the code that uses, for every message, the optimal single-message state at $s=P$, carried to that message by the unitaries of step (i). The same holds for $B'$, by exchanging the roles of the two bases.

\emph{Numerical evaluation.} To evaluate Eq.~\eqref{eq:seqoned} we write $\lambda\Pi_i+\mu\Pi'_j=\vartheta M_{ij}+\tfrac\delta2(\Pi_i-\Pi'_j)$, solve the constraint for $\vartheta$ at fixed $\delta$, which is possible because $\Tr(\rho M_{ij})$ increases strictly with $\vartheta$, and minimize over $\delta$.

\emph{Qubits, Eq.~\eqref{eq:seqone}.} For $d=2$ we take $B=\sigma_z$ and $B'=\sigma_x$; by step (i) it suffices to treat the message $(0,0)$. For $\rho=\tfrac12(\id+\bm r\cdot\bm\sigma)$ the success of the message is $s=\tfrac14(2+r_z+r_x)$, and $\Irr_B=h_2(\tfrac{1+r_z}2)-h_2(\tfrac{1+|\bm r|}2)$, the entropy of the outcome of $\sigma_z$ minus that of the state. A component $r_y$ does not change $s$ and only increases $|\bm r|$, which increases $\Irr_B$; the minimum therefore has $r_y=0$, as do the states $\rho_{\lambda\mu}$. At fixed $s$, what remains free is how the success is shared between the two decoders. Writing $r_{z,x}=\upsilon/2\pm\epsilon$, with $\upsilon=r_z+r_x=2(2P-1)$, gives $|\bm r|^2=\upsilon^2/2+2\epsilon^2$, so the condition $|\bm r|\le1$ becomes $|\epsilon|\le\omega$; this is Eq.~\eqref{eq:seqone}. The irrealism vanishes if and only if $r_x=r_y=0$, i.e., the state is diagonal in the eigenbasis of $\sigma_z$; with $r_x=0$ the constraint requires $r_z=\upsilon$, which is possible exactly when $\upsilon\le1$, i.e., $P\le3/4$. Beyond $3/4$ the minimum leaves zero with zero slope: the pure state with $(r_z,r_x)=(\cos a,\sin a)$ and $\cos a+\sin a=\upsilon$, for which $a\simeq\upsilon-1=4(P-\tfrac34)$, has $\Irr_B=h_2(\cos^2\tfrac a2)=O\big(a^2\ln a^{-1}\big)$, which bounds the minimum from above. At $P=P_Q$, $\omega=0$ and $|\bm r|=1$: the only feasible state is pure.

\begin{figure}[t]
\centering
\includegraphics[width=\columnwidth]{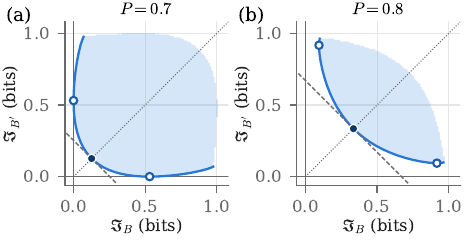}
\caption{Attainable irrealisms of the two decoding observables for one qubit message, with $B=\sigma_z$ and $B'=\sigma_x$, at success probability (a) $P=0.7<P_C$ and (b) $P=0.8>P_C$. Solid curve: states with $r_y=0$, which form the lower-left boundary; the shaded region is filled by sampled states with $r_y\neq0$. Open circles: minimum of $\Irr_B$ or of $\Irr_{B'}$ alone, Eq.~\eqref{eq:seqone}; filled circle: the Gibbs state, which minimizes $\Irr_B+\Irr_{B'}$, Eq.~\eqref{eq:qubit}, at the point where a line of constant sum (dashed) touches the region; dotted: the diagonal $\Irr_B=\Irr_{B'}$. Below $P_C$ the region reaches the axes, so either decoding observable can be an element of reality, but never the origin.}
\label{fig:region}
\end{figure}

\emph{Strict inequality~\eqref{eq:strict}.} At $\epsilon=0$, that is, for equal decoding probabilities, the first argument in Eq.~\eqref{eq:seqone} is $(2+\upsilon)/4=P$ and the second is $\tfrac12+\tfrac{2P-1}{\sqrt2}$, so twice the bracket is the frontier~\eqref{eq:qubit}. Moving to small $\epsilon>0$ lowers the bracket. Its second term is even in $\epsilon$ and does not change to first order, whereas the derivative of the first term at $\epsilon=0$ is $\tfrac12\log_2[(1-P)/P]=\tfrac12\log_2[(2-\upsilon)/(2+\upsilon)]$, negative for $\upsilon>0$, i.e., $P>\tfrac12$. Since $\omega>0$ for $P<P_Q$, small $\epsilon>0$ lies inside the allowed interval and gives a smaller value, which proves Eq.~\eqref{eq:strict}: the minimum for one decoding observable is reached with unequal decoding probabilities, the Gibbs state with equal ones. Figure~\ref{fig:region} shows the two minima in the plane of the two irrealisms.

\section{Excess-irrealism identity}\label{app:lemma}

The identity holds for any weighted sum of irrealisms. Let $\Phi_1,\dots,\Phi_n$ be dephasing maps $\Phi_k(X)=\sum_m\Pi^{(k)}_mX\Pi^{(k)}_m$, let $c_k>0$, and let $F=\sum_kc_k\Irr_{\Phi_k}$, where $\Irr_\Phi(\rho)=S(\Phi(\rho))-S(\rho)$. Suppose that a full-rank state $\rho$ satisfies
\begin{equation}
\sum_kc_k\big[\log\rho-\log\Phi_k(\rho)\big]=\kappa\id+\textstyle\sum_j\zeta_jO_j
\label{eq:closureG}
\end{equation}
for Hermitian $O_j$ and real $\kappa,\zeta_j$ (logarithms in the same base as in $F$; changing the base only rescales $\kappa$ and $\zeta_j$). Then every state $\sigma$ with $\Tr(\sigma O_j)=\Tr(\rho O_j)$ for all $j$ obeys
\begin{equation}
F(\sigma)-F(\rho)=\sum_kc_k\Big[D(\sigma\Vert\rho)-D\big(\Phi_k(\sigma)\Vert\Phi_k(\rho)\big)\Big]\ \ge\ 0 ,
\label{eq:excessG}
\end{equation}
and if equality holds, $\rho^{-1/2}\sigma\rho^{-1/2}$ commutes with all projectors $\Pi^{(k)}_m$, for every $k$. Equations~\eqref{eq:closure} and~\eqref{eq:excess} are the case $\Phi_1=\Phi_B$, $\Phi_2=\Phi_{B'}$, $c_1=c_2=1$, with the single constraint $O_1=M$.

Work with natural logarithms; the statement in bits follows by dividing by $\ln2$. For a dephasing $\Phi$ and a full-rank $\rho$, $\Phi(\rho)$ is full rank, $\Phi$ is self-adjoint with respect to $\langle X,Y\rangle=\Tr X^\dagger Y$, and $\log\Phi(\rho)$ is invariant under $\Phi$. Hence, for every state $\tau$,
\begin{equation}
D(\Phi(\tau)\Vert\Phi(\rho))=-S(\Phi(\tau))-\Tr[\tau\log\Phi(\rho)],
\end{equation}
and therefore
\begin{equation}
\begin{aligned}
D(\tau\Vert\rho)&-D(\Phi(\tau)\Vert\Phi(\rho))\\&=\Irr_\Phi(\tau)-\Tr\big[\tau\big(\log\rho-\log\Phi(\rho)\big)\big].
\end{aligned}
\end{equation}
Multiplying by $c_k$ and summing over $k$ gives
\begin{equation}
\sum_kc_k\Big[D(\tau\Vert\rho)-D\big(\Phi_k(\tau)\Vert\Phi_k(\rho)\big)\Big]
=F(\tau)-\Tr(\tau G),
\end{equation}
where $G=\sum_kc_k[\log\rho-\log\Phi_k(\rho)]=\kappa\id+\sum_j\zeta_jO_j$ by
Eq.~\eqref{eq:closureG}. For $\tau=\rho$ the left-hand side vanishes, so $F(\rho)=\Tr(\rho G)$. For $\tau=\sigma$, the constraints give $\Tr(\sigma G)=\Tr(\rho G)$, which proves Eq.~\eqref{eq:excessG}. Nonnegativity is the monotonicity of relative entropy~\cite{Lindblad1975,Uhlmann1977}; all terms are finite because $\rho$ and $\Phi_k(\rho)$ are full rank, so the identity holds for every competitor $\sigma$ satisfying the constraints, including rank-deficient ones. The operator $G$ is the gradient of $F$ at $\rho$ (up to a factor $1/\ln2$ in bits), so the right-hand side of Eq.~\eqref{eq:excessG} is the Bregman remainder $F(\sigma)-F(\rho)-\Tr[(\sigma-\rho)\nabla F(\rho)]$: closure says that this gradient is a combination of the identity and the constrained observables, and Eq.~\eqref{eq:excessG} then turns stationarity into global optimality. Closure is thus the Karush--Kuhn--Tucker condition $\nabla F(\rho)\in\mathrm{span}\{\id,O_j\}$ for the convex functional $F$, a sum of relative entropies to dephased states of the form minimized in numerical key-rate computations~\cite{Coles2016,Winick2018}; Eq.~\eqref{eq:excessG}, an identity of the Pythagorean type familiar from information projections~\cite{Csiszar1975}, adds the explicit form of the excess, which gives uniqueness through Petz's theorem. If equality holds, each gap vanishes. The second argument $\rho$ is faithful while the competitor $\sigma$ may be rank deficient, and for this case the equality condition in the monotonicity of relative entropy~\cite{Petz1986,JencovaPetz2006,Hiai2011} states that $\sigma$ is recovered by the Petz map, $\sigma=\rho^{1/2}\,\Phi_k^\dagger\big[Y_k\big]\,\rho^{1/2}$ with $Y_k=\Phi_k(\rho)^{-1/2}\,\Phi_k(\sigma)\,\Phi_k(\rho)^{-1/2}$. Since $\Phi_k^\dagger=\Phi_k$ maps every operator into the commutant of $\{\Pi^{(k)}_m\}$, $\rho^{-1/2}\sigma\rho^{-1/2}=\Phi_k(Y_k)$ commutes with every $\Pi^{(k)}_m$, which proves the equality condition.

\section{Gibbs frontier}\label{app:gibbs}

\emph{Spectrum.} Choose the phase of $\ket{b'_j}$ so that $\langle b_i|b'_j\rangle=d^{-1/2}$. In the orthonormal basis $\{\ket{b_i},\ket{b_\perp}\}$ of $\mathrm{span}\{\ket{b_i},\ket{b'_j}\}$, with $\ket{b_\perp}\propto\ket{b'_j}-d^{-1/2}\ket{b_i}$, the restriction of $M=\Pi_i+\Pi'_j$ to this span has trace $2$ and determinant $1-1/d$. Its nonzero eigenvalues are therefore $\lambda_\pm=1\pm d^{-1/2}$, and $M$ vanishes on the $(d-2)$-dimensional orthogonal complement. This gives $Z_d(\theta)=\Tr e^{\theta M}$ in Eq.~\eqref{eq:gibbs}. The eigenvector for $\lambda_+$ is $\ket{\psi_{ij}}\propto\ket{b_i}+e^{-\mathrm{i}\arg\langle b_i|b'_j\rangle}\ket{b'_j}$, the standard optimal state~\cite{Tavakoli2015}.

\emph{Equal decoding probabilities.} The Householder reflection $R$ that exchanges $\ket{b_i}$ and $e^{-\mathrm{i}\arg\langle b_i|b'_j\rangle}\ket{b'_j}$, and acts as the identity on the complement, is unitary and satisfies $RMR^\dagger=M$ and $R\Pi_iR^\dagger=\Pi'_j$. Hence $\Tr(\rho_\theta \Pi_i)=\Tr(\rho_\theta \Pi'_j)=\tfrac12\Tr(\rho_\theta M)=\tfrac12\,d\ln Z_d/d\theta$, which is Eq.~\eqref{eq:Ptheta}. Moreover, $dP/d\theta=\tfrac12\mathrm{Var}_{\rho_\theta}(M)>0$ because $M\not\propto\id$. As $\theta\to-\infty$ the state concentrates on the smallest eigenvalue of $M$, which is $0$ for $d>2$ and $\lambda_-$ for $d=2$. This gives the stated range.

\emph{Dephased state.} Because $M$ has at most three distinct eigenvalues, $e^{\theta M}=\xi_0\id+\xi_1M+\xi_2M^2$. Using $M^2=\Pi_i+\Pi'_j+\Pi_i\Pi'_j+\Pi'_j\Pi_i$ and $\Pi_i\ket{b_k}=0$ for $k\neq i$, we obtain $\bra{b_k}M\ket{b_k}=\bra{b_k}M^2\ket{b_k}=|\langle b_k|b'_j\rangle|^2=1/d$. All diagonal elements with $k\neq i$ are therefore equal, which gives Eq.~\eqref{eq:deph}. Thus
\begin{equation}
\ln\Phi_B(\rho_\theta)=\ln\frac{1-P}{d-1}\,\id+\alpha \Pi_i,\qquad\alpha=\ln\frac{(d-1)P}{1-P},
\end{equation}
and likewise for $B'$. With $\ln\rho_\theta=\theta M-\ln Z_d\,\id$, condition~\eqref{eq:closure} holds, in natural logarithms, with $M=M_{ij}$ and $\zeta=2\theta-\alpha$.

\emph{Value.} We have $S(\Phi_B(\rho_\theta))=S(\Phi_{B'}(\rho_\theta))=h_2(P)+(1-P)\log_2(d-1)$ and $S(\rho_\theta)=[\ln Z_d-\theta\Tr(\rho_\theta M)]/\ln2=\log_2Z_d-2\theta P/\ln2$, which give Eq.~\eqref{eq:frontier}.

\emph{Single message.} By Eq.~\eqref{eq:excess}, $g_d(s)\equiv\min\{\Irr_B(\sigma)+\Irr_{B'}(\sigma):\Tr(\sigma M_{ij})=2s\}$ is attained by $\rho_{ij}(\theta)$ with $P(\theta)=s$, for every $s$ in the open range of Eq.~\eqref{eq:Ptheta}, including $s<1/d$ ($\theta<0$). The resulting function is the same for every pair $(i,j)$. Uniqueness follows from the equality condition. The operator $Y=\rho^{-1/2}\sigma\rho^{-1/2}$ commutes with every $\Pi_k$ and every $\Pi'_k$, so $Y=\sum_ky_k\Pi_k=\sum_ly'_l\Pi'_l$. Then $y_k\langle b_k|b'_l\rangle=y'_l\langle b_k|b'_l\rangle$ with $\langle b_k|b'_l\rangle\neq0$, so $Y=y\,\id$ and $\sigma=\rho$. At $s=P_Q$ the only feasible state is $\proj{\psi_{ij}}$, because $\Tr(\sigma M)=\lambda_+$ forces $\sigma$ onto the top eigenvector; no continuity argument is needed at that endpoint.

\emph{Slope and strict convexity.} Differentiating Eq.~\eqref{eq:frontier} along $\theta$ and using $Z_d'/Z_d=2P$ gives Eq.~\eqref{eq:slope}. Differentiating again,
\begin{equation}
\frac{d}{d\theta}\Big[2\theta-\alpha\Big]=2-\frac{P'(\theta)}{P(1-P)}=2-\frac{\mathrm{Var}(M)}{2P(1-P)} .
\end{equation}
Now $\mathrm{Var}(M)=\mathrm{Var}(\Pi_i)+\mathrm{Var}(\Pi'_j)+2\,\mathrm{Cov}$, with $\mathrm{Var}(\Pi_i)=\mathrm{Var}(\Pi'_j)=P(1-P)$, because $\Pi_i$ and $\Pi'_j$ are projectors with mean $P$, and $\mathrm{Cov}=\mathrm{Re}\Tr[\rho(\Pi_i-P\id)(\Pi'_j-P\id)]$. Because $\rho_\theta>0$, $\langle X,Y\rangle_\rho=\Tr(\rho X^\dagger Y)$ is an inner product, and the Cauchy--Schwarz inequality gives $\mathrm{Cov}\le P(1-P)$. Equality would require $\Pi_i-P\id=\xi(\Pi'_j-P\id)$ for a real $\xi$, which is impossible for two noncommuting rank-one projectors. Hence $\mathrm{Var}(M)<4P(1-P)$, the slope is strictly increasing in $\theta$, and since $P'(\theta)>0$, $g_d$ is strictly convex on the open range of $s$. The slope vanishes at $\theta=0$, where $P=1/d$, so $g_d$ increases on $[1/d,P_Q]$, and the constraints $P_{\rm PM}=P$ and $P_{\rm PM}\ge P$ give the same frontier.

\emph{Comparison with the exclusion bound.} At $\theta=0$, $\rho_\theta=\id/d$ and $\mathrm{Var}(M)=\Tr(M^2)/d-(2/d)^2=2(d-1)/d^2$, since $\Tr M^2=2+2/d$. Hence $dP/d\theta=(d-1)/d^2$, and differentiating Eq.~\eqref{eq:slope} gives $d^2\cC^{(d)}_{\min}/dP^2=(2/\ln2)[2\,d\theta/dP-d\alpha/dP]=(2/\ln2)\,d^2/(d-1)$ at $P=1/d$. Since $K_d''(q)=1/[\ln2\,q(1-q)]$, $2K_d''(1/d)$ has the same value. Both functions vanish with zero slope at $P=1/d$, so their difference is at least of third order. It is in fact of fourth order, and the exact minimum lies strictly above the bound for $P>1/d$. Since $dK_d/dP=\alpha/\ln2$, Eq.~\eqref{eq:slope} gives
\begin{equation}
\frac{d}{dP}\big[\cC^{(d)}_{\min}-2K_d(P)\big]=\frac4{\ln2}(\theta-\alpha).
\end{equation}
With $\varsigma=1/\sqrt d$, we have $Z_dP=e^\theta[\cosh(\varsigma\theta)+\varsigma\sinh(\varsigma\theta)]$ and $Z_d(1-P)=d-2+e^\theta[\cosh(\varsigma\theta)-\varsigma\sinh(\varsigma\theta)]$, so $\theta>\alpha$ is equivalent to $\Xi(\theta)>0$, where
\begin{equation}
\begin{aligned}
\Xi(\theta)={}&d-2+e^\theta\big[\cosh(\varsigma\theta)-\varsigma\sinh(\varsigma\theta)\big]\\&-(d-1)\big[\cosh(\varsigma\theta)+\varsigma\sinh(\varsigma\theta)\big]=\sum_{n\ge0}\Xi_n\frac{\theta^n}{n!}.
\end{aligned}
\end{equation}
Using $e^\theta[\cosh(\varsigma\theta)-\varsigma\sinh(\varsigma\theta)]=\tfrac12[(1-\varsigma)e^{(1+\varsigma)\theta}+(1+\varsigma)e^{(1-\varsigma)\theta}]$ and $d-1=(1-\varsigma^2)/\varsigma^2$,
\begin{equation}
\begin{aligned}
\Xi_n=\frac{1-\varsigma^2}2\Big[&(1+\varsigma)^{n-1}+(1-\varsigma)^{n-1}\\&-\varsigma^{n-2}\big(1+\varsigma +(-1)^n(1-\varsigma)\big)\Big],
\end{aligned}
\end{equation}
for $n\ge1$, so that $\Xi_1=\Xi_2=0$; the constant $d-2$ enters only $\Xi_0$, and $\Xi_0=\Xi(0)=0$. For $n\ge3$, $(1+\varsigma)^{n-1}+(1-\varsigma)^{n-1}>2$ while the subtracted term is $2\varsigma^{n-2}$ or $2\varsigma^{n-1}$, both smaller than $2$ because $\varsigma <1$; hence $\Xi_n>0$. Therefore $\Xi(\theta)>0$ for $\theta>0$, the difference $\cC^{(d)}_{\min}-2K_d$ is strictly increasing on $[1/d,P_Q]$ and positive for $P>1/d$. Moreover, $e^{\theta-\alpha}=1+\Xi(\theta)/\{(d-1)[\cosh(\varsigma\theta)+\varsigma\sinh(\varsigma\theta)]\}$, so $\theta-\alpha=O(\theta^3)$; since $P-1/d$ is of first order in $\theta$, the difference is $O\big((P-1/d)^4\big)$. Explicitly, $\theta-\alpha=\Xi_3\theta^3/[6(d-1)]+O(\theta^4)$ with $\Xi_3=1-\varsigma^2$, and $P-1/d=(d-1)\theta/d^2+O(\theta^2)$, so that integrating gives
\begin{equation}
\begin{aligned}
\cC^{(d)}_{\min}(P)-2K_d(P)={}&\frac{d^5}{6\ln2\,(d-1)^3}\Big(P-\frac1d\Big)^4\\&+O\Big(\big(P-\tfrac1d\big)^5\Big),
\end{aligned}
\end{equation}
with coefficient $7.69$ for $d=2$, $7.30$ for $d=3$, and $18.7$ for $d=7$.

\emph{Full code.} The function $g_d$ is defined on the closed range $[s_-,P_Q]$ of $s$, with $s_-=0$ for $d>2$ and $s_-=\tfrac12-\tfrac1{2\sqrt2}$ for $d=2$; at $s_-$ the feasible states are those supported on the eigenspace of the smallest eigenvalue of $M_{ij}$, and at $P_Q$ the only feasible state is $\proj{\psi_{ij}}$. It is the value function of a convex functional under a linear constraint, since $\Irr_B+\Irr_{B'}$ is convex in the state by the joint convexity of relative entropy, and it is therefore convex on the closed range. A convex function that is strictly convex on the interior of an interval is strictly convex on the whole closed interval, because an affine piece on a segment with an endpoint at the boundary would be affine on the interior part of that segment. For any code, $\cC_{\rm PM}\ge d^{-2}\sum_{ij}g_d(s_{ij})\ge g_d(P_{\rm PM})$ by Jensen's inequality, with equality for the Gibbs code. Strict convexity forces $s_{ij}=P$ for all $(i,j)$ in any optimal code, and single-message uniqueness then fixes every $\rho_{ij}$. The Gibbs state has $d-2$ equal eigenvalues $1/Z_d$ and two further ones, $e^{\theta\lambda_\pm}/Z_d$, whereas the isotropic family has $d-1$ equal eigenvalues. For $d>2$ the two coincide only if $e^{\theta\lambda_-}=1$, i.e., $\theta=0$, or in the pure limit, which excludes the isotropic family for every $P\in(1/d,P_Q)$. For $d=2$ the Gibbs states are themselves of the isotropic form.

\emph{Window, Eq.~\eqref{eq:window}.} For each message, $\Irr_B(\rho)+\Irr_{B'}(\rho)=S(\Phi_B(\rho))+S(\Phi_{B'}(\rho))-2S(\rho)\le H_B+H_{B'}$, where $H_B$ is the Shannon entropy of the outcome distribution of $B$. Among $d$-outcome distributions with a given entry $p$, the entropy is maximized by spreading $1-p$ uniformly, so $H_B\le f_d(p_{ij})$ and $H_{B'}\le f_d(q_{ij})$. Since $f_d''(p)=-1/[\ln2\,p(1-p)]<0$, Jensen's inequality over the $2d^2$ probabilities $p_{ij},q_{ij}$, whose mean is $P_{\rm PM}$, gives $\cC_{\rm PM}\le2f_d(P_{\rm PM})$. With $S(\rho_\theta)=\log_2Z_d-2\theta P/\ln2$, Eq.~\eqref{eq:frontier} reads $\cC^{(d)}_{\min}(P)=2f_d(P)-2S(\rho_\theta)$. Equality in the upper bound requires pure states whose two outcome distributions have the intended entry $P$ and are uniform elsewhere. At $P_Q$ the standard optimal states have this property, by Eq.~\eqref{eq:deph} in the limit $\theta\to\infty$. For $d=2$ and message $(0,0)$, the pure state with Bloch vector $(2P-1,\pm[1-2(2P-1)^2]^{1/2},2P-1)$ has it for every $P\in[\tfrac12,P_Q]$, and the symmetry unitaries of Appendix~\ref{app:seqone} (here Pauli operators) give the other messages. Finally, $dS(\rho_\theta)/d\theta=-\theta\,\mathrm{Var}_{\rho_\theta}(M)/\ln2<0$ for $\theta>0$, and $S(\rho_\theta)\to0$ as $\theta\to\infty$.

\emph{Qubits.} For $d=2$, $Z_2=2e^\theta\cosh(\theta/\sqrt2)$ and $\rho_\theta=\tfrac12[\id+\tanh(\theta/\sqrt2)\,\hat n\cdot\bm\sigma]$, so $S(\rho_\theta)=h_2(\tfrac12+\tfrac{2P-1}{\sqrt2})$. Since $\log_2(d-1)=0$, this yields Eq.~\eqref{eq:qubit}.

\section{Asymptotics}\label{app:asym}

With $L=\ln d$, the endpoints of the frontier behave as
\begin{align}
\cC_Q(d)&=\log_2d+2-\frac{\log_2d}{\sqrt d}-\frac{2}{d\ln2}+O(d^{-3/2}),\label{eq:IQas}\\
\cC_C(d)&=\log_2d-\frac{\theta_C^2+L-1}{d\ln2}+O\Big(\frac{L^4}{d^2}\Big),\label{eq:ICas}
\end{align}
where $\theta_C$, defined by $P(\theta_C)=P_C(d)$, diverges as $\theta_C\simeq2L-\ln(4L)$. The leading term, $\log_2d-\cC_C\simeq4L^2/(d\ln2)$, is accurate only at very large $d$, so Eq.~\eqref{eq:ICas} should be used with $\theta_C$ itself. We derive both expansions below.

\emph{Quantum endpoint.} We have $h_2(P_Q)=1-\tfrac{1}{2d\ln2}+O(d^{-2})$ and $\log_2(d-1)=\log_2d-\tfrac1{d\ln2}+O(d^{-2})$. Substituting into $\cC_Q=2[h_2(P_Q)+\tfrac12(1-d^{-1/2})\log_2(d-1)]$ gives Eq.~\eqref{eq:IQas}; the first neglected term is $+\,d^{-3/2}/\ln2$.

\emph{Classical endpoint.} Write $w=(d-2)/Z_d$ for the weight of the kernel of $M$ and $\delta=\tanh(\theta/\sqrt d)$. The two nonzero-mode weights are $\tfrac12(1-w)(1\pm\delta)$, and exactly
\begin{align}
P&=\tfrac12(1-w)(1+\delta/\sqrt d),\\
\ln2\,S(\rho_\theta)&=h(w)+(1-w)h\big(\tfrac{1+\delta}2\big)+w\ln(d-2),
\end{align}
with $h$ the binary entropy in nats. At $P=P_C$ the first gives $w=(\delta/\sqrt d-1/d)/(1+\delta/\sqrt d)$, so that, with $\delta/\sqrt d=\theta/d+O(\theta^3/d^2)$, $\nu\equiv dw=\theta-1+O(L^3/d)$. The definition of $w$ gives $2e^\theta\cosh(\theta/\sqrt d)=(d-2)(1-w)/w$, hence
\begin{equation}
2L=\nu+1+\ln(2\nu)+O(L^3/d),
\end{equation}
whose solution, with $\theta_C=\nu+1+O(L^3/d)$, is $\theta_C=2L-\ln(4L)+(1+\ln4L)/(2L)+O(\ln^2L/L^2)$. Next, $\ln2\,(\log_2d-\cC_C)=L-2h(P_C)-2(1-P_C)\ln(d-1)+2\ln2\,S(\rho_\theta)$. Inserting $h(P_C)=\ln2+O(d^{-2})$, $(1-P_C)\ln(d-1)=\tfrac L2-\tfrac{L+1}{2d}+O(L/d^2)$, $h(w)=-w\ln w+w+O(w^2)$ and $h(\tfrac{1+\delta}2)=\ln2-\tfrac{\theta^2}{2d}+O(\theta^4/d^2)$, and multiplying by $d$,
\begin{equation}
\begin{aligned}
d\ln2\,(\log_2d-\cC_C)={}&L+1+4\nu L-2\nu\ln(2\nu)\\&+2\nu-\theta^2+O(L^4/d).
\end{aligned}
\end{equation}
With $\theta=\nu+1$ and $4\nu L=2\nu[\nu+1+\ln(2\nu)]$, up to $O(L^4/d)$, the right-hand side reduces to $\theta_C^2+L-1$, which is Eq.~\eqref{eq:ICas}. Its expansion, $4L^2-4L\ln L+(1-8\ln2)L+O(\ln^2L)$, shows why the two leading terms alone are inaccurate at moderate $d$. A high-precision evaluation of the frontier confirms Eq.~\eqref{eq:ICas} (relative error of $\log_2d-\cC_C$ below $10^{-16}$ at $d=10^{20}$). Combining Eqs.~\eqref{eq:IQas} and~\eqref{eq:ICas} gives $\cC_Q-\cC_C=2-\log_2d/\sqrt d+O(L^2/d)\to2$. The expansion of $\theta_C$ converges slowly: the residual of its three-term form, $\theta_C-[2L-\ln(4L)+(1+\ln4L)/(2L)]$, is $0.012$ at $d=10^3$ and $0.0016$ at $d=10^{20}$.

\emph{Nonmonotonic gap.} The gap $\cC_Q-\cC_C$ is not monotonic in $d$. The inset of Fig.~\ref{fig:dim}(c) shows a local maximum, $1.98$, at $d=35$, followed by a decrease to a minimum, $1.855$, at $d=936$; it then returns to two bits from below, as Eqs.~\eqref{eq:IQas} and~\eqref{eq:ICas} imply, and in every case we computed (all $d<3000$ and $d=10^k$ up to $10^{12}$) it stays below two bits.

\emph{Slopes.} The slope of the frontier at $P_C$, $(2/\ln2)[2\theta_C-\ln(d+1)]$, grows only logarithmically with $d$, whereas the mean slope $(\cC_Q-\cC_C)/(P_Q-P_C)\simeq4\sqrt d$, because $P_Q-P_C\simeq1/(2\sqrt d)$; at $d=10^6$ they are about $98$ and $4\times10^3$ bits per unit of success probability.

\section{Entanglement-assisted codes}\label{app:EA}

\emph{Trade-off~\eqref{eq:tradeoff}.} (a) \emph{Correlations.} A traceless observable $A$ with $\|A\|\le1$ has the form $\bm a\cdot\bm\sigma$ with $|\bm a|\le1$, and likewise $A'=\bm a'\cdot\bm\sigma$, so that $\cS=\bm a\cdot(\bm u+\bm v)+\bm a'\cdot(\bm u-\bm v)\le|\bm u+\bm v|+|\bm u-\bm v|\le2(|\bm u|^2+|\bm v|^2)^{1/2}$ by the Cauchy--Schwarz and parallelogram identities, the step that leads to the criterion of Ref.~\cite{Horodecki1995}. Moreover, $|\bm u|=\langle\hat{\bm u}\cdot\bm\sigma\otimes\sigma_z\rangle\le1$, and likewise $|\bm v|\le1$.

(b) \emph{Exclusion.} Apply Eq.~\eqref{eq:exclusion2} with Alice measuring $\hat{\bm v}\cdot\bm\sigma$. The right-hand side of Eq.~\eqref{eq:exclusion2} equals $1-H(b'|a)$, and $\Pr(a=b')=(1+|\bm v|)/2$. Hence
\begin{equation}
\Irr_B(\rho)\ge1-H(b'|a)\ge1-H(a\oplus b')=\varphi(|\bm v|),
\end{equation}
and likewise $\Irr_{B'}(\rho)\ge\varphi(|\bm u|)$. The second inequality holds because $H(b'|a)=H(a\oplus b'|a)\le H(a\oplus b')$, and $a\oplus b'$ is a binary variable with bias $|\bm v|$. For a qubit, $\Phi_B(\sigma)=\tfrac12(\sigma+B\sigma B)$ is an equal mixture of two unitarily related states, so $S(\Phi_B(\rho))\le S(\rho)+1$ and $\Irr_B\le1$, and likewise for $B'$. Since $\Irr_B,\Irr_{B'}\in[0,1]$, the increasing function $\varphi^{-1}$ can be applied to the two bounds, giving $|\bm v|\le\varphi^{-1}(\Irr_B)$ and $|\bm u|\le\varphi^{-1}(\Irr_{B'})$; together with (a) this yields Eq.~\eqref{eq:tradeoff} for a qubit on Alice's side and traceless observables.

(c) \emph{Any system on Alice's side.} A joint Naimark dilation of Alice's two measurements, with a common ancilla in a fixed pure state on her side, leaves the irrealisms and $\cS$ unchanged and makes $A,A'$ projective. Jordan's lemma then provides a decomposition $\mathcal H_{\cA}=\bigoplus_k\mathcal H_k$, with $\dim\mathcal H_k\le2$, that reduces both observables; we split commuting pairs, including blocks on which one observable acts as $\pm\id$, into one-dimensional blocks, so that in every two-dimensional block both spectral projections of each observable have rank one and both observables are traceless. The block dephasing $\Lambda(\rho)=\sum_k(\Gamma_k\otimes\id)\rho(\Gamma_k\otimes\id)$, with $\Gamma_k$ the projector onto $\mathcal H_k$, leaves $\cS$ unchanged and cannot increase $\Irr_B$ or $\Irr_{B'}$ (data processing, since $\Lambda$ commutes with $\Phi_B$ and $\Phi_{B'}$). For $\Lambda(\rho)=\bigoplus_kq_k\rho_k$ one has $\Irr_B=\sum_kq_k\Irr_B(\rho_k)$, likewise for $B'$, and $\cS=\sum_kq_k\cS_k$. Let $F(x,y)=[\varphi^{-1}(x)]^2+[\varphi^{-1}(y)]^2$ on $[0,1]^2$. The function $\chi$ maps $[0,1]$ onto $[0,1]$, with $\chi(0)=0$ and $\chi(1)=1$, and is convex and increasing, so its inverse $\chi^{-1}(x)=[\varphi^{-1}(x)]^2$ is concave and increasing on $[0,1]$, and $F$ is concave and nondecreasing in each argument. In a two-dimensional block, (a) and (b) give $F(\Irr_B(\rho_k),\Irr_{B'}(\rho_k))\ge\cS_k^2/4$. In a one-dimensional block Alice's observables are numbers $a,a'\in\{\pm1\}$ and $\rho_k$ is a state of Bob's qubit with Bloch vector $\bm r$, so that $\cS_k=a(r_z+r_x)+a'(r_z-r_x)$ and $|\cS_k|\le2\max\{|r_z|,|r_x|\}$. There $\Irr_B(\rho_k)=\varphi(|\bm r|)-\varphi(|r_z|)$, and since $\chi$ is convex with $\chi(0)=0$, hence superadditive, $\varphi(|\bm r|)\ge\chi(r_z^2+r_x^2)\ge\varphi(|r_z|)+\varphi(|r_x|)$; thus $\Irr_B(\rho_k)\ge\varphi(|r_x|)$ and, likewise, $\Irr_{B'}(\rho_k)\ge\varphi(|r_z|)$, so that $F\ge r_x^2+r_z^2\ge\cS_k^2/4$. Combining the blocks,
\begin{equation}
\begin{aligned}
F(\Irr_B,\Irr_{B'})&\ge F\Big(\sum_kq_k\Irr_B(\rho_k),\sum_kq_k\Irr_{B'}(\rho_k)\Big)\\
&\ge\sum_kq_k\frac{\cS_k^2}{4}\ge\frac{\cS^2}4,
\end{aligned}
\end{equation}
by monotonicity, concavity of $F$, and convexity of $x^2$. This is Eq.~\eqref{eq:tradeoff} for any system on Alice's side and any two-outcome measurements of hers.

(d) \emph{Extremal states.} Using $\sigma_y\otimes\sigma_y=-(\sigma_z\otimes\sigma_z)(\sigma_x\otimes\sigma_x)$, the state $\rho_{t_z,t_x}$ assigns weight $\tfrac14(1+s_zt_z)(1+s_xt_x)$ to the Bell state with $\sigma_z\otimes\sigma_z=s_z$ and $\sigma_x\otimes\sigma_x=s_x$. It is therefore a state, with $S(\rho)=h_2(\tfrac{1+t_z}2)+h_2(\tfrac{1+t_x}2)$. An unrevealed measurement of $\sigma_z$ on Bob's side leaves $\tfrac14(\id+t_z\sigma_z\otimes\sigma_z)$, whose entropy is $1+h_2(\tfrac{1+t_z}2)$. Hence $\Irr_B=\varphi(t_x)$ and, likewise, $\Irr_{B'}=\varphi(t_z)$. Here $\bm u=t_z\hat z$ and $\bm v=t_x\hat x$, and Alice's observables along $\bm u\pm\bm v$ give $\cS=2(t_z^2+t_x^2)^{1/2}$. For $t_z=t_x=t$ the Bell weights are proportional to $e^{\pm2\sqrt2\beta}$ or $1$, with $t=\tanh(\sqrt2\beta)$, i.e., $\rho_{t,t}\propto e^{\beta W}$.

\emph{Threshold~\eqref{eq:threshold}.} Since $\varphi^{-1}(\Irr_{B'})\le1$, Eq.~\eqref{eq:tradeoff} gives $[\varphi^{-1}(\Irr_B)]^2\ge\cS^2/4-1$, which is Eq.~\eqref{eq:threshold} for any system on Alice's side; for $|\cS|\le2$ the bound is trivial. If $B$ is an element of reality, $\Irr_B=0$ and Eq.~\eqref{eq:threshold} forces $\cS\le2$.

\emph{Total irrealism, Eqs.~\eqref{eq:EAsv} and~\eqref{eq:EAcost}.} The first inequality in Eq.~\eqref{eq:EAsv} is Eq.~\eqref{eq:memory} with $d_\cB=2$. For the second, write $\rho=\tfrac14\big(\id+\bm r_\cA\cdot\bm\sigma\otimes\id+\id\otimes\bm r_\cB\cdot\bm\sigma+\sum_{kl}T_{kl}\,\sigma_k\otimes\sigma_l\big)$. Local unitaries act as $T\mapsto R_ATR_B^{\top}$ with $R_A,R_B\in SO(3)$ and leave $S(\cB|\cA)$ unchanged; we choose them so that $T=\mathrm{diag}(\tau_x,\tau_y,\tau_z)$, where the $|\tau_k|$ are the singular values of $T$. The twirl $\Lambda(\rho)=\tfrac14\sum_{k=0}^3(\sigma_k\otimes\sigma_k)\rho(\sigma_k\otimes\sigma_k)$, with $\sigma_0=\id$, leaves the diagonal of $T$ unchanged, while each $k\neq0$ reverses two components of $\bm r_\cA$ and of $\bm r_\cB$, so that the average removes them. The conditional entropy is concave, by strong subadditivity~\cite{Lieb1973}, and invariant under local unitaries, hence $S(\cB|\cA)_\rho\le S(\cB|\cA)_{\Lambda(\rho)}$. As in (d), $\Lambda(\rho)$ assigns a probability $\lambda(s_z,s_x)$ to the Bell state with $\sigma_z\otimes\sigma_z=s_z$ and $\sigma_x\otimes\sigma_x=s_x$, and $\tau_z=\langle s_z\rangle$, $\tau_x=\langle s_x\rangle$, $\tau_y=-\langle s_zs_x\rangle$. Any two of the variables $s_z$, $s_x$, $s_zs_x$ determine the third, so $\lambda$ is a joint distribution of two bits with biases $|\tau_k|$ and $|\tau_l|$, for any chosen pair $k\neq l$. Since $\Tr_\cB\Lambda(\rho)=\id/2$, $S(\cB|\cA)_{\Lambda(\rho)}=H(\lambda)-1$, and subadditivity gives $H(\lambda)\le h_2(\tfrac{1+|\tau_k|}2)+h_2(\tfrac{1+|\tau_l|}2)$. The choice of the two largest $|\tau_k|$ yields $1-S(\cB|\cA)_\rho\ge\varphi(t_{(1)})+\varphi(t_{(2)})$. For $\rho_{t_z,t_x}$ the twirl acts trivially and $\lambda$ is a product distribution, so with (d) every inequality in Eq.~\eqref{eq:EAsv} is an equality.

For Eq.~\eqref{eq:EAcost}, note that $\bm u=T\hat z$ and $\bm v=T\hat x$. Thus $|\bm u|^2+|\bm v|^2$ is the trace of $T^{\top}T$ restricted to the plane of $\hat z$ and $\hat x$, which cannot exceed the sum $t_{(1)}^2+t_{(2)}^2$ of its two largest eigenvalues, and (a) gives $\cS^2/4\le t_{(1)}^2+t_{(2)}^2$ for a qubit on Alice's side and traceless observables. The convexity and monotonicity of $\chi$ give $\varphi(t_{(1)})+\varphi(t_{(2)})\ge2\chi\big(\tfrac{t_{(1)}^2+t_{(2)}^2}2\big)\ge2\chi(\cS^2/8)$. The same steps applied to (b) give $\cC_{\rm EA}\ge\varphi(|\bm u|)+\varphi(|\bm v|)\ge2\chi(\cS^2/8)$. Equality in Eq.~\eqref{eq:EAcost} holds for $\rho_{t,t}$ with $t=\cS/(2\sqrt2)$. With Alice's measurements along $\bm u\pm\bm v\propto\hat z\pm\hat x$, Bob's Bloch vector conditioned on her outcome $a$ is $(-1)^a\,T^{\top}\hat{\bm a}$, with $\hat{\bm a}$ the unit vector of her measurement and $T=\mathrm{diag}(t,-t^2,t)$, which gives the steered states quoted in Sec.~\ref{sec:EA}.

\emph{Arbitrary system on Alice's side, Eq.~\eqref{eq:EAcost}.} Let $\Psi(\cS)=2\varphi\big(\max\{\cS,0\}/(2\sqrt2)\big)$, a convex nondecreasing function of $\cS$, being a convex nondecreasing function of a convex nondecreasing one. As in (c) above, a Naimark dilation and the block dephasing $\Lambda$ leave $\cS$ unchanged, cannot increase $\cC_{\rm EA}$, and give $\cC_{\rm EA}(\Lambda(\rho))=\sum_kq_k\cC_{\rm EA}(\rho_k)$ and $\cS=\sum_kq_k\cS_k$, with $\cS_k\le2\sqrt2$. In a two-dimensional block both projective observables of Alice have rank-one spectral projectors, so they are traceless, and the qubit result gives $\cC_{\rm EA}(\rho_k)\ge \Psi(\cS_k)$. In a one-dimensional block Alice's observables are numbers $a,a'\in\{\pm1\}$, and $\rho_k$ is a state of Bob's qubit with Bloch vector $\bm r$. Then $\cS_k=a(r_z+r_x)+a'(r_z-r_x)\le2m$, with $m=\max\{|r_z|,|r_x|\}$, and $\cC_{\rm EA}(\rho_k)=2\varphi(|\bm r|)-\varphi(|r_z|)-\varphi(|r_x|)$. Let $t=\min\{|r_z|,|r_x|\}$. Setting $r_y=0$ can only lower $\cC_{\rm EA}(\rho_k)$, and with $r=(m^2+t^2)^{1/2}$,
\begin{equation}
\frac{\partial\cC_{\rm EA}(\rho_k)}{\partial t}=\frac1{\ln2}\Big[\frac{2t\,\mathrm{artanh}\,r}{r}-\mathrm{artanh}\,t\Big]\ \ge\ 0,
\end{equation}
because $\mathrm{artanh}(x)/x$ is increasing and $r\ge t$. Hence $\cC_{\rm EA}(\rho_k)\ge\varphi(m)$, the value at $t=0$. The series~\eqref{eq:phi} gives $\varphi(m)\ge2\varphi(m/\sqrt2)$, term by term, since $2^{1-k}\le1$; and $2\varphi(m/\sqrt2)\ge \Psi(\cS_k)$ because $\cS_k\le2m$. Thus $\cC_{\rm EA}(\rho_k)\ge \Psi(\cS_k)$ in every block, and Jensen's inequality gives $\cC_{\rm EA}(\rho)\ge\sum_kq_k\Psi(\cS_k)\ge \Psi(\cS)$, which is Eq.~\eqref{eq:EAcost} for any quantum system on Alice's side and any two-outcome measurements of hers.

\emph{Steering decomposition~\eqref{eq:steer} and bound~\eqref{eq:steerbound}.} Let $\mathcal M_z(\rho)=\sum_a\proj a\otimes\Tr_\cA[(E^z_a\otimes\id)\rho]$, where $\{E^z_a\}$ is Alice's POVM. Since $\mathcal M_z(\rho)=\sum_ap_a\proj a\otimes\rho_\cB^a$ is classical--quantum, the argument of Sec.~\ref{sec:PM} gives $\Irr_B(\mathcal M_z(\rho))=\sum_ap_a\Irr_B(\rho^a_\cB)$, which proves Eq.~\eqref{eq:steer}. Since $\mathcal M_z$ acts on Alice's side, $\mathcal M_z\circ\Phi_B=\Phi_B\circ\mathcal M_z$, and data processing gives $\Delta^z_B\ge0$.

For Eq.~\eqref{eq:steerbound}, average Eq.~\eqref{eq:steer} over $z$ with weight $\tfrac12$. The steered ensemble with weights $\pi_{(a,z)}=p(a|z)/2$ is a PM code. The probability that the outcome of $B$ ($B'$) equals $a$ ($a\oplus z$) is the probability that his answer is correct for $y=0$ ($y=1$), so $\sum\pi\,s=P_{\rm EA}$. Then $\sum\pi\,(\Irr_B+\Irr_{B'})(\rho_\cB^{a|z})\ge\sum\pi\,g_2(s)\ge g_2(P_{\rm EA})$, by the message-independent single-message bound of Appendix~\ref{app:gibbs} (Pauli operators, which map messages into one another, as in Appendix~\ref{app:seqone}) and Jensen's inequality with weights. Finally, $g_2=\cC^{(2)}_{\min}$ on $[1/2,P_Q]$.

\emph{Reality gain, Eq.~\eqref{eq:CMI}.} Let $\ket{\psi}_{\cA\cB E}$ purify $\rho_{\cA\cB}$, and let Alice's measurement be projective with rank-one projectors $\proj{e_a}$. The purification identity gives $\Irr_B(\rho_{\cA\cB})=H(B|E)$. Conditioned on outcome $a$, the state of $\cB E$ is pure, $\ket{\psi_a}\propto(\bra{e_a}\otimes\id)\ket\psi$, so $E$ purifies $\rho_\cB^{a|z}$ and $\Irr_B(\rho^{a|z}_\cB)=H(B|E)_{\psi_a}$. By the same argument, $\Irr_B(\mathcal M_z(\rho_{\cA\cB}))=\sum_ap(a|z)H(B|E)_{\psi_a}=H(B|E,a)$, where the conditional entropy is evaluated on the state of the outcomes of $B$ and $a$ and the system $E$. Subtracting gives Eq.~\eqref{eq:CMI}. For a general measurement $\{E^z_a\}$, a Naimark isometry maps $\cA$ into a remainder $\cA'$ and an outcome register; the state of $\cA'\cB E$ and the register is pure, and the irrealism of the classical--quantum state $\mathcal M_z(\rho_{\cA\cB})$ is $H(B|\cA'E,a)$, so that $\Delta^z_B=H(B|E)-H(B|\cA'E,a)=I(B{:}\cA'a|E)\ge I(B{:}a|E)$, by the chain rule. Equality holds for rank-one projective measurements, for which $\cA'$ is trivial.

\emph{Certification of realism-based nonlocality, Eq.~\eqref{eq:rbncert}.} Let Alice's measurements be rank-one projective, $\{\proj{e^z_a}\}_{a=0,1}$, so that her system is a qubit, and let Bob decode with $B=\sigma_z$ and $B'=\sigma_x$. The proof combines an identity with three inequalities.

(i) \emph{Success probability of the steered code.} Bob's answer is correct if and only if the outcome of $B$ equals $a$ ($y=0$), or the outcome of $B'$ equals $a\oplus z$ ($y=1$). With $\pi^B_{az}=\Tr(\rho^{a|z}_\cB\Pi_a)$ and $\pi^{B'}_{az}=\Tr(\rho^{a|z}_\cB\Pi'_{a\oplus z})$ the corresponding probabilities in the steered state, and $z$ and $y$ uniform,
\begin{equation}
P_{\rm EA}=\sum_{z,a}\frac{p(a|z)}{4}\big(\pi^B_{az}+\pi^{B'}_{az}\big).
\label{eq:Psteer}
\end{equation}

(ii) \emph{Exact accounting.} Averaging Eq.~\eqref{eq:steer} and its counterpart for $B'$ over $z$ gives the identity $\cC_{\rm EA}=\cC^{\rm st}_{\rm PM}+\bar\Delta$, with $\cC^{\rm st}_{\rm PM}=\tfrac12\sum_{z,a}p(a|z)\big[\Irr_B(\rho^{a|z}_\cB)+\Irr_{B'}(\rho^{a|z}_\cB)\big]$. For a rank-one projective measurement, $\mathcal M_z(\rho_{\cA\cB})$ and $\Phi_{A_z}(\rho_{\cA\cB})=\sum_a\proj{e^z_a}\otimes p(a|z)\rho^{a|z}_\cB$ differ by the local unitary $\ket{e^z_a}\mapsto\ket a$ on Alice's side, which leaves $\Irr_B$ unchanged. Hence $\Delta^z_B=\eta_{B|A_z}(\rho_{\cA\cB})$, Eq.~\eqref{eq:rbn}, and likewise $\Delta^z_{B'}=\eta_{B'|A_z}(\rho_{\cA\cB})$.

(iii) \emph{Upper bound on $\bar\Delta$.} Each $A_z$ is a nondegenerate observable of Alice and $B$, $B'$ are nondegenerate observables of Bob, so each of the four contexts enters the maximum that defines $\eta_{\max}(\rho_{\cA\cB})$, and $\bar\Delta=\tfrac12\sum_z(\eta_{B|A_z}+\eta_{B'|A_z})\le2\eta_{\max}(\rho_{\cA\cB})$.

(iv) \emph{Upper bound on the steered code.} For a qubit state $\sigma$, $\Irr_B(\sigma)=S(\Phi_B(\sigma))-S(\sigma)\le S(\Phi_B(\sigma))=h_2(\Tr\sigma\Pi_k)$ for either $k$, since $h_2(p)=h_2(1-p)$. Hence $\Irr_B(\rho^{a|z}_\cB)\le h_2(\pi^B_{az})$ and $\Irr_{B'}(\rho^{a|z}_\cB)\le h_2(\pi^{B'}_{az})$. The weights $p(a|z)/4$ of the eight probabilities $\pi^B_{az}$, $\pi^{B'}_{az}$ sum to one, and the concavity of $h_2$ together with Eq.~\eqref{eq:Psteer} gives
\begin{equation}
\cC^{\rm st}_{\rm PM}\le2\sum_{z,a}\frac{p(a|z)}4\big[h_2(\pi^B_{az})+h_2(\pi^{B'}_{az})\big]\le2h_2(P_{\rm EA}),
\end{equation}
which is Eq.~\eqref{eq:window} for the steered code with nonuniform weights. Neither $P_{\rm EA}\ge\tfrac12$ nor the monotonicity of $h_2$ is used.

(v) \emph{Lower bound on $\cC_{\rm EA}$.} Equation~\eqref{eq:EAcost}, which holds for any system on Alice's side, gives $\cC_{\rm EA}\ge2\varphi(\cS/2\sqrt2)$ with $\cS/2\sqrt2=\sqrt2(2P_{\rm EA}-1)$; for $P_{\rm EA}<\tfrac12$ the bound holds with $|2P_{\rm EA}-1|$ after relabeling Alice's outcomes, and $\varphi$ is even.

Combining (ii), (iv) and (v), $\bar\Delta=\cC_{\rm EA}-\cC^{\rm st}_{\rm PM}\ge2\varphi\big(\sqrt2(2P_{\rm EA}-1)\big)-2h_2(P_{\rm EA})$, and (iii) gives Eq.~\eqref{eq:rbncert}. The bound uses only the success probability. Its right-hand side vanishes at $P_{\rm EA}=0.81452$, where the largest total irrealism of a steered code, $2h_2(P_{\rm EA})$, equals the EA frontier; below this value the bound is trivial. Two steps lose tightness: (iv) replaces the steered code by its worst case, and (iii) bounds an average over four contexts by the maximum over all contexts. The restriction to rank-one projective measurements enters only in (ii): for a coarser measurement $\mathcal M_z$ is a coarse-graining of $\Phi_A$ for a refining observable $A$, data processing gives $\Delta^z_B\ge\eta_{B|A}$, and the inequality points the wrong way for (iii).

\section{Arbitrary qubit decoders}\label{app:decoders}

Let $B$ and $B'$ be rank-one projective qubit observables whose Bloch axes $\hat n$ and $\hat n'$ make an angle $\gamma\in(0,\pi)$, with outcomes labeled so that $\gamma$ is the angle between the Bloch vectors of $\ket{b_k}$ and $\ket{b'_k}$. The replacement $\gamma\to\pi-\gamma$ only relabels the outcomes of $B'$. In this appendix $y$ and $y_i$ denote Bloch lengths, not Bob's question.

\emph{Single message.} The decoding operator is $M_{ij}=\Pi_i+\Pi'_j=\id+\bm m_{ij}\cdot\bm\sigma$, with $\bm m_{ij}=\tfrac12[(-1)^i\hat n+(-1)^j\hat n']$ and $|\bm m_{ij}|=c$, where $c=\cos(\gamma/2)$ for $i=j$ and $c=\sin(\gamma/2)$ for $i\neq j$. A dephased qubit state lies in the span of $\id$ and $\Pi_i$, and the Gibbs state $e^{\theta M_{ij}}/\Tr e^{\theta M_{ij}}$ has Bloch vector $y\,\hat m_{ij}$, with $y=\tanh(\theta c)$, whose projections on $(-1)^i\hat n$ and $(-1)^j\hat n'$ both equal $cy$. Its two decoding probabilities are therefore equal, the closure property~\eqref{eq:closure} holds with $M=M_{ij}$, and by Eq.~\eqref{eq:excess} the Gibbs state is the unique minimizer of $\Irr_B+\Irr_{B'}$ at fixed $s_{ij}=\tfrac12(1+cy)$. For a qubit, $\Irr_B=\Info-\Info_B$ with $\Info=\varphi(|\bm r|)$ and $\Info_B=\varphi(|\bm r\cdot\hat n|)$, and likewise for $B'$, so the minimum is $\ell_c(x)=2[\varphi(x/c)-\varphi(x)]$, with $x=cy\in[0,c]$. It is increasing and strictly convex, since $\ell_c''(x)=(2/\ln2)[(c^2-x^2)^{-1}-(1-x^2)^{-1}]>0$. Since $\varphi$ is strictly increasing, $\ell_c(x)$ strictly decreases with $c$ at fixed $x>0$, that is, at fixed success $s_{ij}>\tfrac12$.

\emph{Full code.} The messages form two classes of two, with $c_1=\cos(\gamma/2)$ and $c_2=\sin(\gamma/2)$, so that $c_1^2+c_2^2=1$. By strict convexity the optimal code assigns the same success to the messages of a class, and
\begin{equation}
\cC_{\min}(P\,|\,\gamma)=\min_{y_1,y_2}\sum_{i=1,2}\big[\varphi(y_i)-\varphi(c_iy_i)\big],
\label{eq:anydec}
\end{equation}
over $0\le y_i\le1$ with $c_1y_1+c_2y_2=2(2P-1)$, a one-parameter convex problem, defined for $P\le\tfrac12+\tfrac14(c_1+c_2)$. For $\gamma=\pi/2$ the minimum is at $y_1=y_2$ and Eq.~\eqref{eq:anydec} reduces to Eq.~\eqref{eq:qubit}. For $B'=B$ the code that sends $\ket{b_i}$ reaches $P=3/4$ with zero irrealism.

\emph{Mutually unbiased decoders as a function of $P$.} Write $u_i=c_iy_i$, $v_i=c_i^2$, and $f=\ln2\,\varphi$, so that $f'=\mathrm{artanh}$. In nats the cost is $\mathcal G(u_1,v_1)+\mathcal G(u_2,v_2)$, with $\mathcal G(u,v)=f(u/\sqrt v)-f(u)$, under the constraints $u_1+u_2=2X$, $X=2P-1$, and $v_1+v_2=1$. Mutually unbiased decoders with their optimal states are the symmetric point $(u_i,v_i)=(X,\tfrac12)$. Rotating the decoders and reallocating the success moves the two points to $(X\pm\delta,\tfrac12\pm\epsilon)$, and the cost changes by $(\delta,\epsilon)J(\delta,\epsilon)^{\top}+O(\delta^4+\epsilon^4)$, with $J$ the Hessian of $\mathcal G$ at $(X,\tfrac12)$. With $y=\sqrt2X$ the Bloch length of the optimal state for mutually unbiased decoders, $T_y=y^2/(1-y^2)$, and $A_y=y\,\mathrm{artanh}\,y$,
\begin{equation}
\begin{aligned}
\mathcal G_{uu}&=\frac2{1-y^2}-\frac2{2-y^2},\\
\mathcal G_{uv}&=-\frac{\sqrt2(T_y+A_y)}{y},\qquad \mathcal G_{vv}=T_y+3A_y,
\end{aligned}
\end{equation}
and
\begin{equation}
\det J=-\frac{2\,Q(y)}{(1-y^2)(2-y^2)},
\end{equation}
with
\begin{equation}
\begin{aligned}
Q(y)={}&(1-y^2)(2-y^2)\,\mathrm{artanh}^2y\\&-y(2y^2-1)\,\mathrm{artanh}\,y+y^2 .
\end{aligned}
\label{eq:Q}
\end{equation}
Since $\mathcal G_{uu}>0$, mutually unbiased decoders are a strict local minimum over decoders if $Q(y)<0$. If $Q(y)>0$ they are not a local minimum, and a direction of decrease necessarily has $\epsilon\neq0$: there are decoders that are not mutually unbiased and require strictly less total irrealism. A strict local minimum in $(\delta,\epsilon)$ is a strict local minimum of $\gamma\mapsto\cC_{\min}(P|\gamma)$, because for fixed $\epsilon$ the problem in $\delta$ is strictly convex, so its minimizer is unique and tends to $0$ with $\epsilon$. At $P=P_*$ the Hessian is singular, and nothing is claimed there.

As a quadratic in $a=\mathrm{artanh}\,y$, $Q$ has discriminant $y^2(8y^2-7)$, so $Q>0$ for $y^2<7/8$. For $y^2\ge7/8$, $Q<0$ if and only if $a_-<\mathrm{artanh}\,y<a_+$, with roots
\begin{equation}
\begin{aligned}
a_-&=\frac{2y}{2y^2-1+\sqrt{8y^2-7}},\\
a_+&=\frac{y\,\big(2y^2-1+\sqrt{8y^2-7}\big)}{2(1-y^2)(2-y^2)} .
\end{aligned}
\end{equation}
Since $2/a_-=2y-1/y+(8-7/y^2)^{1/2}$ increases with $y$, $a_-$ is strictly decreasing, from $8y/3=2.494$ at $y^2=7/8$, where $\mathrm{artanh}\,y=1.700$, to $1$ at $y=1$, while $\mathrm{artanh}\,y$ increases without bound; the two cross exactly once. Moreover, $a_+\ge y/[3(1-y^2)]$ because $2y^2-1\ge3/4$ and $2-y^2\le9/8$. For $y^2\ge8/9$, $\mathrm{artanh}\,y\le y/\sqrt{1-y^2}\le y/[3(1-y^2)]$, from $\ln z\le(z-1)/\sqrt z$; for $7/8\le y^2\le8/9$, $\mathrm{artanh}\,y\le\ln(3+2\sqrt2)=1.763<2.494\le y/[3(1-y^2)]$. Hence $\mathrm{artanh}\,y<a_+$ throughout, and $Q(y)<0$ if and only if $y>y_*$, where $y_*$ is the root of $\mathrm{artanh}\,y=a_-(y)$, i.e.,
\begin{equation}
y_*\simeq0.9415,\qquad P_*=\frac12+\frac{y_*}{2\sqrt2}\simeq0.8329
\label{eq:Pstar}
\end{equation}
For $\tfrac12<P<P_*$ mutually unbiased decoders therefore do not minimize the total irrealism, and for $P_*<P<P_Q$ they are a strict local minimum over decoders; at $P_Q$ they are the only decoders that reach the success probability~\cite{Farkas2019}. The difference below $P_*$ is substantial. Minimizing Eq.~\eqref{eq:anydec} numerically over $\gamma$ gives $0.134$, $0.371$, and $0.589$ bits at $P=0.76$, $0.78$, and $0.80$, against $0.463$, $0.557$, and $0.670$ bits for mutually unbiased decoders; at $P=0.83$ the two differ by about $10^{-3}$ bits. Above $P_*$ the same minimization, on a grid of $2\times10^4$ angles in $(0,\pi/2]$, places the minimum at $\gamma=\pi/2$ for every $P$ tested, so that mutually unbiased decoders are, numerically, the global minimum for $P_*<P\le P_Q$. Because Eq.~\eqref{eq:anydec} is invariant under $\gamma\to\pi-\gamma$, which exchanges $c_1$ and $c_2$, the point $\gamma=\pi/2$ is always critical.

\end{document}